\documentclass[a4paper,11pt]{article}
\usepackage{jcappub}
\usepackage{lineno}
\usepackage{subcaption}
\nolinenumbers 

\arxivnumber{2506.09319}


\title{Machine Learning Left-Right Breaking from Gravitational Waves}

\author[a]{W. Searle,}
\author[a]{C. Bal\'azs,}
\author[b,c,d]{Y. Xiao,}
\author[b]{and Y. Zhang}
\affiliation[a]{School of Physics and Astronomy, Monash University, Melbourne 3800 Victoria, Australia}
\affiliation[b]{School of Physics, Henan Normal University, Xinxiang 453007, P. R. China}
\affiliation[c]{Institute of Theoretical Physics, Chinese Academy of Sciences, Beijing 100190, China}
\affiliation[d]{School of Physical Sciences, University of Chinese Academy of Sciences,  Beijing 100049, China}

\emailAdd{william.searle@monash.edu}

\abstract{ 
First-order phase transitions in the early universe can generate stochastic gravitational waves (GWs), offering a unique probe of high-scale particle physics.  The Left-Right Symmetric Model (LRSM), which restores parity symmetry at high energies and naturally incorporates the seesaw mechanism, allows for such transitions --- particularly during the spontaneous breaking of $SU(2)_R \times SU(2)_L \times U(1)_{B-L} \to SU(2)_L \times U(1)_Y$.
This initial step, though less studied, is both theoretically motivated and potentially observable via GWs.
In this work, we investigate the GW signatures associated with this first-step phase transition in the minimal LRSM.  Due to the complexity and dimensionality of its parameter space, traditional scanning approaches are computationally intensive and inefficient.  To overcome this challenge, we employ a Machine Learning Scan (MLS) strategy, integrated with the high-precision three-dimensional effective field theory framework --- using \texttt{PhaseTracer} as an interface to \texttt{DRalgo} --- to efficiently identify phenomenologically viable regions of the parameter space.  Through successive MLS iterations, we identify a parameter region that yields GW signals detectable at forthcoming gravitational wave observatories, such as BBO and DECIGO.  Additionally, we analyse the evolution of the MLS-recommended parameter space across iterations and perform a sensitivity analysis to identify the most influential parameters in the model.  Our findings underscore both the observational prospects of gravitational waves from LRSM phase transitions and the efficacy of machine learning techniques in probing complex beyond the Standard-Model landscapes.
}

\numberwithin{equation}{section}

\newcommand{\tr}[1]{\mathrm{Tr}(#1)}
\newcommand{\dg}[1]{#1^{\dagger}}
\newcommand{\tl}[1]{\widetilde{#1}}

\begin{document}
\maketitle
\flushbottom

\section{Introduction}

The early universe underwent a variety of complex processes --- such as cosmic inflation, reheating, and Big Bang nucleosynthesis --- that critically shaped its thermal and structural evolution.  Among these, cosmological phase transitions are of particular interest due to their potential to produce observable relics~\cite{PhysRevD.30.272}.  In particular, first-order phase transitions, characterised by the nucleation and expansion of true-vacuum bubbles, can generate a stochastic GW background through bubble collisions, sound waves, and magnetohydrodynamic turbulence~\cite{kosowsky1992gravitational,kosowsky1993gravitational, jinno2017gravitational, jinno2019gravitational, hindmarsh2014gravitational, hindmarsh2015numerical, hindmarsh2018sound, hindmarsh2017shape, hindmarsh2019gravitational, Caldwell:2022qsj, Caprini:2019egz, Athron:2023xlk}.  If such transitions occurred in the early universe, the resulting GW signals may be detectable today.  Several upcoming or proposed GW detectors --- including LISA~\cite{LISA:2017pwj}, Taiji~\cite{gong2015descope}, TianQin~\cite{luo2016tianqin}, DECIGO~\cite{Corbin_2006} and BBO~\cite{Musha:2017usi} --- are expected to reach the required sensitivity to observe these signals, opening a new observational window into particle cosmology.

Within the Standard Model (SM), lattice simulations show that the electroweak symmetry breaking is a smooth crossover rather than a first-order transition, making the generation of observable gravitational waves highly unlikely~\cite{Csikor:1998eu, Kajantie:1996mn}.  This limitation underscores the importance of exploring Beyond the Standard Model (BSM) scenarios that naturally permit first-order phase transitions.  One of the well-motivated candidates is the Left-Right Symmetric Model (LRSM), which extends the SM gauge symmetry to $SU(2)_L \times SU(2)_R \times U(1)_{B-L}$, restoring parity symmetry at high energies and providing a dynamical explanation for its spontaneous violation at low energies~\cite{PhysRevD.10.275, PhysRevD.11.2558, PhysRevD.12.1502}.  Notably, the LRSM incorporates the seesaw mechanism in a natural way, elegantly accounting for the smallness of neutrino masses~\cite{Mohapatra:2005wg, gu2010left, buchmuller2000neutrino}.  Its rich particle content --- including heavy gauge bosons, scalar triplets, and Majorana fermions --- has been extensively studied in collider phenomenology, flavour physics, and neutrinoless double beta decay~\cite{Dev:2016dja, Patra:2012ur, Deppisch:2017vne, Ramsey-Musolf:2020ndm}.  Experimental constraints from LHC searches and low-energy observables place significant bounds on the symmetry-breaking scale and associated new physics~\cite{Bernard:2020cyi, Karmakar:2022iip, Lindner:2016lxq}.

In the LRSM framework, symmetry breaking generally proceeds via a sequence of spontaneous breakings:
\begin{equation}
SU(2)_R \times SU(2)_L \times U(1)_{B - L} \rightarrow SU(2)_L \times U(1)_Y \rightarrow U(1)_{\text{em}},
\end{equation}
each of which can be first-order under appropriate conditions.  While much of the literature has focused on the second step --- namely the electroweak phase transition --- comparatively little attention has been paid to the first step, during which the left-right symmetry is spontaneously broken~\cite{Brdar:2019fur, Karmakar:2023ixo, Li:2020eun}.  This initial transition is both theoretically well-motivated and cosmologically significant.  It marks the energy scale at which parity symmetry is broken in the early universe, potentially sets the stage for neutrino mass generation via the seesaw mechanism, and provides a potential explanation for the origin of baryon asymmetry~\cite{Gu:2019rzu}.  Moreover, if this phase transition is first-order, the resulting stochastic GW background could carry distinctive imprints of the underlying left-right symmetric structure.  Such signals would provide a rare opportunity to probe the existence and nature of parity restoration at high energies, offering direct evidence for one of the most compelling extensions of the SM.

Motivated by these considerations, our study focuses on a detailed exploration of the first step of symmetry breaking in the minimal LRSM.  Specifically, Our aim is to identify regions of parameter space where this transition is first-order and capable of producing GW signals detectable by future experiments.  Accurately determining the nature of such high-scale transitions requires a precise treatment of the finite-temperature effective potential, which suffers from significant theoretical uncertainties in conventional approaches \cite{Athron:2022jyi, Bittar:2025lcr}.  Specifically, due to the contribution of Matsubara zero modes, the finite-temperature effective potential has severe infrared (IR) divergences~\cite{Laine:2016hma}.  To mitigate this issue, the most divergent contributions are often resummed, leading to the so-called Daisy diagrams~\cite{Parwani:1991gq, Dine:1992vs, Arnold:1992rz}.  However, the Daisy resummation blurs the perturbative expansion order and, as shown in Ref.~\cite{Niemi:2020hto}, the resulting phase structure may still significantly deviate from lattice simulation results and even indicate spurious phase transitions.  To address these problems more reliably, the Three-Dimensional Effective Field Theory (3dEFT) framework has been developed~\cite{Croon:2020cgk, Schicho:2022wty, Ekstedt:2022zro, Ekstedt:2024etx, Kainulainen:2019kyp, Niemi:2024axp, Niemi:2024vzw, Gould:2023ovu}.
By integrating out all non-zero Matsubara modes --- except for the scalar zero modes --- and matching the resulting Green's functions with those of the original 4D theory, one obtains a three-dimensional effective theory that features a well-defined perturbative expansion, improved convergence, and reduced dependence on renormalisation scale~\cite{Croon:2020cgk, Schicho:2022wty, Ekstedt:2022zro, Ekstedt:2024etx}.
Importantly, since this effective theory contains no fermionic degrees of freedom, it is well suited for non-perturbative lattice simulations, which can accurately capture IR effects and yield reliable phase transition information.  Previous studies have shown that even perturbative calculations within the 3dEFT framework agree well with non-perturbative lattice results~\cite{Niemi:2024axp, Niemi:2024vzw, Gould:2023ovu}.  Encouraged by this, we adopt the 3dEFT framework to ensure the reliability of our results.

To implement this approach, we use our numerical code \texttt{PhaseTracer}, which enables end-to-end computations from BSM models to gravitational wave signals~\cite{Athron:2020sbe, Athron:2024xrh}.  It also provides an interface with the 3dEFT tool \texttt{DRalgo}~\cite{Ekstedt:2022bff}, allowing users to automatically analyse the phase transition properties associated with the accurate effective potential.  Given the large dimensionality of the LRSM parameter space and the computational cost of high-precision evaluations, we also incorporate machine learning techniques to efficiently guide and optimise the parameter scan.  This strategy enables us to pinpoint viable regions in the parameter space where the first step of symmetry breaking is first-order and leads to potentially observable gravitational wave signals.

The structure of this paper is as follows.  In Section~\ref{sec::Model}, we introduce the model, detailing the particle content and scalar sector. Section~\ref{sec::SymBreak} presents our framework for symmetry-breaking, including vacuum alignment of the three scalar fields, and the calculation of the effective potential and dimensional reduction.  In Section~\ref{sec::GravWaves}, we describe the generation of gravitational waves, covering thermal parameters, transition temperatures, and the fitting formulas used.  Section ~\ref{sec::analysis} outlines the parameter space and methodology for obtaining observable predictions.  Our results are presented in Section~\ref{sec::results}, followed by conclusions in Section~\ref{sec::conclusion}.



\section{The Left-Right Standard Model} \label{sec::Model}


The scalar content of the minimal LRSM can be obtained via a LR `symmetrization' of the SM.  The SM Higgs doublet is extended to a $SU(2)_L \times SU(2)_R$ bi-doublet, $\phi$, belonging to the representation $(2, 2)_0$ of the LR gauge group.  A right-handed Higgs triplet, $\Delta_R$, belonging to the representation $(3, 1)_2$, is incorporated to break the parity symmetry in the first step via a vacuum expectation value (VEV), $v_R$.  Additionally, a left-handed triplet $\Delta_L$, with representation $(1, 3)_2$, is included to ensure that the scalar sector is symmetric under $L \leftrightarrow R$.  Electroweak symmetry breaking occurs when $\phi$ and $\Delta_L$ take the VEVs $v_h$ and $v_L$ respectively.  These fields can be written most easily as:
\begin{equation}
    \phi = \begin{pmatrix} \phi_1^0 & \phi_1^+ \\ \phi_2^{-} & \phi_2^0\end{pmatrix} \text{~ and ~} \Delta_i = \begin{pmatrix} \delta_i^{+}/\sqrt{2} & \delta_i^{++} \\ \delta_i^0 & - \delta_i^{+}/\sqrt{2} \end{pmatrix},
\end{equation}
where $i=L,R$, and we have the corresponding transformation laws:
\begin{equation}
    \phi \rightarrow U_L \phi U_R^{\dagger}, \ \Delta_L \rightarrow U_L \Delta_L U_L^{\dagger}, \ \Delta_R \rightarrow U_R \Delta_R U_R^{\dagger}.
\end{equation}
Here, $U_L \in SU(2)_L$ and $U_R \in SU(2)_R$ belong to the fundamental representation.  Likewise, these fields transform under $U(1)_{B-L}$ as:
\begin{equation}
    \phi \rightarrow \phi, \ \Delta_i \rightarrow e^{i \theta_{B-L}} \Delta_i,
\end{equation}
and under $\mathcal{P}: L \leftrightarrow R$ as:
\begin{equation}
    \mathcal{P}: \phi \rightarrow \phi^{\dagger}, \ \Delta_L \leftrightarrow \Delta_R
    \label{eq:ParityTransform}
\end{equation}
Note that this transformation, $\mathcal{P}: \Delta_L \rightarrow \Delta_R$, is only unique up a complex phase.  This redundancy is accounted for in the gauge degree of freedom provided by $U(1)_{B-L}$. 

Breaking of the $SU(2)_R \times SU(2)_L$ group breaks this discrete $\mathcal{P}$ symmetry.  If both triplets acquire vacuum expectation values, $\langle \Delta_L \rangle, \langle \Delta_R \rangle \ne 0$, we expect the formation of domain walls as the VEV transitions into either $L$ and $R$-breaking configurations over super-horizon scales.  These domain walls can act as an additional source of gravitational waves ~\cite{Borboruah:2022eex}, particularly relevant to the nano-Hertz frequency band.  We assume a parameter space wherein either the $L$ or $R$ triplet acquires a non-zero VEV, but avoid the case where both can.  As such, we do not consider the domain wall contribution to our spectrum.

The $W_{i,\mu}^a$ $SU(2)_i$ bosons, together with the $B_{\mu}$ $U(1)_{B-L}$ boson, couple to the Higgs sector via the following covariant derivative terms:
\begin{equation}
    \mathcal{L}_g = \tr{ \dg{(D_{\mu} \Delta_L )} (D^{\mu} \Delta_L ) }  + \tr{ \dg{(D_{\mu} \Delta_R )} (D^{\mu} \Delta_R ) } + \tr{ \dg{(D_{\mu} \phi )} (D^{\mu} \phi)}.
\end{equation}
The above derivatives are given by:
\begin{equation}
    \begin{split}
    D_{\mu} \Delta_i & = \partial_{\mu} \Delta_i - i g_i [ W_{i, \mu}^a \tau^a , \Delta_i ] - i g_{BL} B_{\mu} \Delta_i, \\
    D_{\mu} \phi & = \partial_{\mu} \phi - i g_L (W_{L, \mu}^a \tau^a \phi - \phi W_{R, \mu}^a \tau^a ).
    \end{split}
\end{equation}
In the first step, we neglect VEVs from both $\Delta_L$ and $\phi$, from which the above terms produce a gauge boson mass spectrum analogous to the SM case, producing two massive $W_R^{\pm}$ bosons, a massive $Z_R$ boson, and the massless $U(1)_{Z}$ boson.  To ensure that the above terms are symmetric under $\mathcal{P}$, we consider only the cases where $g_L = g_R$ \cite{Maiezza:2016ybz}.

We arrange the SM fermions into the following LRSM representations \cite{BhupalDev:2018xya}:
\begin{equation}
    Q_L = \begin{pmatrix} u_L \\ d_L\\ \end{pmatrix}: (3, 2, 1)_{\frac{1}{3}}, \ \ \ Q_R = \begin{pmatrix} u_R \\ d_R\\ \end{pmatrix}: (3, 1, 2)_{\frac{1}{3}};
\end{equation}
\begin{equation}
    \psi_L = \begin{pmatrix} \nu_L \\ e_L\\ \end{pmatrix}: (1, 2, 1)_{-1}, \ \ \ \psi_R = \begin{pmatrix} N_R \\ e_R\\ \end{pmatrix}: (1, 1, 2)_{-1}.
\end{equation}
We have made clear the distinction between quarks and leptons by including representations under $SU(3)_c$.  Note that the right-handed lepton doublet features a right-handed neutrino, $N_R$, not typically present in the SM.  During $SU(2)_R$ breaking, a seesaw mechanism generates a mass for $N_R$ of the order $v_R$, while $\nu_L$ gains a small mass of the order $\mathcal{O}(v_h^2/v_R)$, where $v_h$ is the SM Higgs VEV.  For a sub-eV left-handed neutrino mass, this immediately places the constraint $v_R > v_h^2 \gtrsim 10^4$ GeV.

\paragraph{Yukawa Sector} \label{subsec::yukawa}

The above fermions couple to the Higgs fields via a Yukawa sector that contains both quark-Higgs and lepton-Higgs couplings.  The distinction is raised as the former features no contribution from the triplet fields, and takes the form:
\begin{equation}
    \mathcal{L}_Y \supset \overline{Q}_L^i \left ( \mathcal{F}_{i j} \phi + \mathcal{G}_{ij} \tl{\phi} \right ) Q^j_R + H.c.,
\end{equation}
where $i$ sums over quark flavours, whilst $\mathcal{F}$ and $\mathcal{G}$ are the Yukawa matrices.  This term is responsible for generating the SM quark masses during the EW phase transition.  To ensure the correct neutrino masses are generated during the first step, the lepton-Higgs couplings contain both Dirac and Majorana terms:
\begin{equation}
    \mathcal{L}_Y \supset \overline{\psi}^i_L \left ( f_{ij} \phi + g_{ij} \tl{\phi} \right ) \psi^j_R + i (h_M)_{ij} \left ( {\psi^{i}_L}^T \mathcal{C} \tau_2 \Delta_L \psi^j_L + {\psi^{i}_R}^T \mathcal{C} \tau_2 \Delta_R \psi^j_R \right ) + H.c.,
\end{equation}
where $f$ and $g$ are Yukawa matrices and $\mathcal{C}$ is the charge conjugation operator.  The Majorana matrix is given by $(h_M)_{ij}$.

\paragraph{Higgs Sector} \label{subsec::higgs}

Lastly, full Higgs sector is given in the scalar potential \cite{BhupalDev:2018xya}:
\begin{equation}
    \begin{split}
    V = &  \mu_1^2 \tr { \dg{\phi} \phi}  + \mu_2^2 [ \tr{\tl{\phi} \dg{\phi}} + \tr{ \dg{ \tl{\phi}} \phi}] +  \mu_3^2 [ \tr{ \Delta_L \dg{\Delta_L} + \Delta_R \dg{\Delta_R}}] \\
    & + \lambda_1  \tr{\phi \dg{\phi}} ^2 + \lambda_2 [ \tr{\tl{\phi} \dg{\phi}}^2 + \tr{\dg{\tl{\phi}} \phi}^2 ] + \lambda_3 \tr{ \tl{\phi} \dg{\phi}} \tr{\dg{\tl{\phi}} \phi} \\
    & + \lambda_4 \tr{ \dg{\phi} \phi} [ \tr{ \tl{\phi} \dg{\phi}} + \tr{\dg{\tl{\phi}} \phi}]  + \rho_1 [ \tr{ \Delta_L \dg{\Delta_L}} ^2 + \tr{ \Delta_R \dg{\Delta_R} } ^2 ] \\
    & + \rho_2 [ \tr{\Delta_L \Delta_L} \tr{\dg{\Delta_L} \dg{\Delta_L} } + \tr{\Delta_R \Delta_R} \tr{\dg{\Delta_R} \dg{\Delta_R} }] + \rho_3 [ \tr{ \Delta_L \dg{\Delta_L}} \tr{ \Delta_R \dg{\Delta_R}} ] \\
    & + \rho_4 [ \tr{ \Delta_L \Delta_L } \tr{ \dg{\Delta_R} \dg{\Delta_R}} + \tr{\dg{\Delta_L} \dg{\Delta_L}} \tr{ \Delta_R \Delta_R} ] \\
    & + \alpha_1 \tr{\phi \dg{\phi}} [ \tr{\Delta_L \dg{\Delta_L}} + \tr{ \Delta_R \dg{\Delta_R}}] + \alpha_2 [ \tr{ \phi \dg{\tl{\phi}}} \tr{ \Delta_R \dg{\Delta_R}} + \tr{ \dg{\phi} \tl{\phi}} \tr{\Delta_L \dg{\Delta_L}}] \\
    & + \alpha_2^* [ \tr{ \dg{\phi} \tl{\phi}} \tr{ \Delta_R \dg{\Delta_R}} + \tr{ \phi \dg{\tl{\phi}}} \tr{\Delta_L \dg{\Delta_L}}] + \alpha_3[ \tr{\phi \dg{\phi} \Delta_L \dg{\Delta_L} } + \tr {\dg{ \phi} \phi \Delta_R \dg{\Delta_R}}] \\ 
    & + \beta_1 [\tr{ \phi \Delta_R \dg{\phi} \dg{\Delta_L}} + \tr{ \dg{\phi} \Delta_L \phi \dg{\Delta_R}} ] + \beta_2 [\tr{\tl{\phi} \Delta_R \dg{\phi} \dg{\Delta_L}} + \tr{ \dg{\tl{\phi}} \Delta_L \phi \dg{\Delta_R}} ] \\
    & + \beta_3 [ \tr{\phi \Delta_R \dg{\tl{\phi}} \dg{\Delta_L}} + \tr{ \dg{\phi} \Delta_L \tl{\phi} \dg{\Delta_R }}],
    \end{split}
\end{equation}
where $\tl{\phi} = \sigma_2 \phi^* \dg{\sigma_2} = \sigma_2 \phi^* \sigma_2$ forms an additional LR bi-doublet. 

\section{Symmetry Breaking} \label{sec::SymBreak}

As mentioned, breaking the discrete symmetry $\mathcal{P}$ allows for creation of domain walls in this model.  To avoid the formation of such structures, we assume that only the right-handed triplet has a VEV in the first step of the transition.  The EW phase transition is then initiated by the VEVs in $\phi$ and $\Delta_L$.  Thus, we have:
\begin{equation}
    SU(2)_R \times SU(2)_L \times U(1)_{B-L} \xrightarrow{\langle \Delta_R \rangle \ne 0 } SU(2)_L \times U(1)_Y \xrightarrow{\langle \phi \rangle, \langle \Delta_L \rangle \ne 0 } U(1)_{em}.
\end{equation}
In Appendix~\ref{App::VEV}, we show that under the assumptions of no CP violation and no $\beta$ couplings, we can assume the following vacuum alignment:
\begin{equation}
    \langle \phi \rangle = \frac{1}{\sqrt{2}} \begin{pmatrix} \kappa_1 & 0 \\ 0 & \kappa_2 \end{pmatrix}, \  \langle \Delta_L \rangle = 0, \ \langle \Delta_R \rangle = \frac{1}{\sqrt{2}} \begin{pmatrix} 0 & 0 \\ v_R & 0 \end{pmatrix}.
    \label{eq:VEValign}
\end{equation}
Indeed, assuming $\alpha_2$ is real and $\beta_i = 0$ allows one to show $v_L= 0$, avoiding domain walls entirely for this region of the parameter space.  The only dynamical field with which we are concerned for the first step of the transition is thus $v_R$.  We will re-parameterise the fields $\kappa_1$ and $\kappa_2$ using 
\begin{equation}
    \kappa_+^2 = \kappa_1^2 + \kappa_2^2 \text{ and } r = \kappa_2/\kappa_1.
\end{equation}
The latter is assumed to be small, and the former is matched to the SM Higgs VEV $v_h \sim 246$ GeV.  With this VEV alignment, we have the following symmetry-breaking conditions:
\begin{equation}
    \frac{\partial V}{\partial \kappa_1} = \frac{\partial V}{\partial \kappa_2} = \frac{\partial V}{\partial v_R} = 0.
\end{equation}
These are utilised to re-derive the Lagrangian masses $\mu_i^2$ in terms of $\kappa_{1,2}$ and $v_R$ (equivalently $\kappa_+$, $r$, and $v_R$).

\paragraph{Fermion Masses} \label{subsec::fermionmass}

The only relevant fermion mass at the symmetry-breaking scale is the right-handed neutrino mass, which would then provide a contribution to the one-loop effective potential.  However, we derive the effective potential within the 3dEFT framework.  Such an approach yields a purely bosonic theory, so there is no such inclusion to the effective potential from a right-neutrino contribution.  This effectively removes a degree of freedom from our study, as we no longer need to consider the Majorana coupling for this field (which would otherwise appear in the effective potential).

However, Yukawa couplings still appear in the thermal effective field theory approach.  Such terms arise due to the presence of fermion loops in the construction of matching relations during the dimensional reduction.  Of these terms, we include the top and bottom quark Yukawa couplings, with all others being parametrically smaller.  Taking only the third generation of quarks into account, we assume the quark Yukawa matrices are of the form $\mathcal{F}_{ij} = \text{diag}(0, 0, y_1)$ and $\mathcal{G}_{ij} = \text{diag}(0, 0, y_2)$ ~\cite{Karmakar:2023ixo}.  This leads to top and bottom masses in the flavour basis of:
\begin{equation}
    m_t = (y_1 \kappa_1 + y_2 \kappa_2) / \sqrt{2}, \quad m_b = (y_1 \kappa_2 + y_2 \kappa_1)/ \sqrt{2}.
\end{equation}
Assuming $V^{CKM}_{33} \sim 1$, we obtain the Yukawa couplings $y_1$ and $y_2$ in terms of the observed top and bottom masses as \cite{Karmakar:2023ixo}:
\begin{equation}
    y_1 = \frac{\sqrt{2} \sqrt{1 + r^2}}{\kappa_+ ( 1 - r^2)} ( m_t - r m_b), \quad y_2 = \frac{\sqrt{2}\sqrt{1 + r^2}}{\kappa_+ ( 1 - r^2 )} (m_b - r m_t).
\end{equation}
For small $r$, these reduce to the familiar values upon setting $\kappa_+ = v_h$:
\begin{equation}
    y_1 = \sqrt{2} m_t/v_h, \quad y_2 = \sqrt{2} m_b / v_h.
\end{equation}

\paragraph{Gauge Boson Masses} \label{subsec::vectormass}

Whilst fermion masses do not appear in the effective theory, both gauge and scalar eigenstates arise in the one-loop potential.  Indeed in many cases, the contribution from gauge boson eigenstates contributes the $g \phi^3$ cubic term responsible for generating a barrier between the two minima.  From the VEV alignments above, the charged gauge bosons have the mass matrix:
\begin{equation}
    \frac{g_L^2}{8} \begin{pmatrix}  W_L^i \\ W_R^i \end{pmatrix}^T \begin{pmatrix} \kappa_+^2 & - 2 \kappa_1 \kappa_2 \\ - 2 \kappa_1 \kappa_2 & \kappa_+^2 + v_R^2 \end{pmatrix} \begin{pmatrix}  W_L^i \\ W_R^i \end{pmatrix},
\end{equation}
Diagonalisation leads to the following expressions for $m_{W}^2$:
\begin{equation}
    m_{W, \pm}^2 = \frac{g_L^2}{8} \bigg ( v_R^2 + \kappa_+^2 \pm \sqrt{ v_R^4 + 4 \kappa_1^2 \kappa_2^2 } \bigg ).
\end{equation}
At the symmetry-breaking scale, $v_R^2 \gg \kappa_+^2$, the different signs in the above expression lead to masses $m_{W, +}^2 = g_L^2 v_R^2/4 \sim v_R^2$ and $m_{W, -}^2 \sim \kappa_+^2$.  The former is identified with the right-handed $W$ bosons, whilst the latter with the left-handed $W$ boson.

Likewise, the neutral gauge bosons mix according to the matrix:
\begin{equation}
    \frac{g_L^2}{8} \begin{pmatrix} W_L^3 \\ W_R^3 \\ B \end{pmatrix}^T \begin{pmatrix}  \kappa_+^2 & -  \kappa_+^2 &  0 \\ -  \kappa_+^2 &  \kappa_+^2 + 4  v_R^2 & - 4 \chi v_R^2 \\ 0 & - 4 \chi v_R^2 & 4 \chi^2 v_R^2 \end{pmatrix} \begin{pmatrix} W_L^3 \\ W_R^3 \\ B \end{pmatrix},
\end{equation}
where $\chi = g_{BL}/g_L$.  Diagonalisation yields a null eigenvalue, which we associate with the $U(1)_{B-L}$ hyper-photon.  We then have for the $Z$ boson masses:
\begin{equation}
    m_{Z,\pm}^2 = \frac{1}{8} \bigg ( g_L^2 \kappa_+^2 + 2 (g_{BL}^2 + g_L^2 ) v_R^2 \pm \sqrt{ 4 g_{BL}^4 v_R^4 + 4 g_{BL}^2 g_L^2 v_R^2 ( 2 v_R^2 - \kappa_+^2 ) + g_L^4 ( 4 v_R^4 + \kappa_+^4 ) } \bigg ).
\end{equation}
-Here, once again, $m_{Z, +}$ is associated with the right-handed $Z$ boson, while $m_{Z, -}$ is the SM $Z$-boson mass.  During the first transition, we assume $\kappa_+ \rightarrow 0$, yielding the gauge boson contributions:
\begin{equation}
    2: \frac{1}{2} (g_{BL}^2 + g_L^2) v_R^2, \quad 4: \frac{1}{4} g_L^2 v_R^2.
\end{equation}
We have listed each eigenstate by its multiplicity. 

\paragraph{Scalar Masses} \label{subsec::scalarmass}

In Appendix~\ref{App::Eigens}, we derive the propagating scalar eigenstates together with their masses for the full VEV alignment Eq.\eqref{eq:VEValign}.  Here, we focus on the case for $\kappa_1 = \kappa_2 = 0$, and provide the eigenstates as they appear in the effective potential.  Applying this limit to the expressions found in Appendix~\ref{App::Eigens}, we obtain:
\begin{align}
    4 & : \mu_1^2 + (\frac{1}{2} \alpha_1 + \frac{1}{4} \alpha_3) v_R^2 + \frac{1}{4} \sqrt{ 16 \alpha_2^2 v_R^4 + \alpha_3 v_R^4 - 64 \alpha_2 \mu^2_2 v_R^2 + 64 \mu_2^4} \label{eq:mass1} \\
    4 & : \mu_1^2 + (\frac{1}{2} \alpha_1 + \frac{1}{4} \alpha_3) v_R^2 - \frac{1}{4} \sqrt{ 16 \alpha_2^2 v_R^4 + \alpha_3 v_R^4 - 64 \alpha_2 \mu^2_2 v_R^2 + 64 \mu_2^4} \label{eq:mass2} \\
    6 & : \mu_3^2 + \frac{1}{2} \rho_3 v_R^2 \label{eq:mass3} \\
    3 & : \mu_3^2 + \rho_1 v_R^2 \label{eq:mass4} \\
    2 & : \mu_3^2 + (3 \rho_1 + 2 \rho_2 ) v_R^2 \label{eq:mass5} \\
    1 & : \mu_3^2 + 3 \rho_1 v_R^2 \label{eq:mass6}
\end{align}
where we have again listed these eigenstates by their multiplicities.  These results are exact; perturbative diagonalisation is only employed for the full scalar spectrum where $\kappa_+ \ne 0$.

Before proceeding, we provide the physical SM Higgs mass, given by Eq.\eqref{eq:PhysHiggsMass}:
\begin{equation}
    m^2_h = \bigg( 2 \lambda_1 - \frac{\alpha_1}{2 \rho_1} \bigg ) \kappa_1^2.
\end{equation}
This was derived in the limit $\kappa_1 \gg \kappa_2$, however restoring powers of $\kappa_2$ is done using the replacement $\kappa_1 \rightarrow \kappa_+$.  We can then compare this to the known SM Higgs mass, $m^2_h = 2 \lambda_{SM} v_h^2$.  As we have $\kappa_+ = v_h$, we can use these two results to constrain $\lambda_1$ in terms of the SM quartic coupling via:
\begin{equation}
    \lambda_1 = \lambda_{SM} + \frac{\alpha_1}{4 \rho_1}.
    \label{eq::lambda1constraint}
\end{equation}

\paragraph{Effective Potential} \label{sec::EffPot} 

In this work, the effective potential is generated within the 3dEFT framework~\cite{Ginsparg:1980ef, Appelquist:1981vg, Nadkarni:1982kb, Farakos:1994kx, Braaten:1995cm, Kajantie:1995dw}.  Among other advantages this approach addresses the slow convergence of conventional finite temperature perturbation theory by constructing a dimensionally-reduced effective field theory, which we refer to as the 3dEFT.  It can be considered a systemic approach to include thermal resummation in the context of perturbation theory~\cite{Ekstedt:2022bff}.  Within this framework, the one loop potential has the following form:
\begin{equation}
    V_{\text{eff}, 3} = V_{0, 3} + V_{1, 3},
\end{equation}
where the subscript 3 denotes quantities belonging to the effective theory. 

$V_{0,3}$ is akin to the conventional tree-level potential with all quantities replaced by their 3dEFT counterparts.  For the full VEV alignment ~Eq.\eqref{eq:VEValign}, the tree-level potential is:
\begin{align*}
V_0 (\kappa_1, \kappa_2, v_R) = & \frac{\mu_1^2}{2} \kappa_1^2 + \frac{\mu_1^1}{2} \kappa_2^2  + 2 \mu_2^2 \kappa_1 \kappa_2  + \frac{\mu_3^2}{2} v_R^2 + \frac{1}{4} \alpha_1 ( \kappa_1^2 + \kappa_2^2) v_R^2 + \alpha_2 \kappa_1 \kappa_2 v_R^2 + \frac{1}{4} \alpha_3 \kappa_2^2 v_R^2 \\ 
& + \frac{1}{4} \lambda_1 ( \kappa_1^2 + \kappa_2^2)^2  + ( 2 \lambda_2 + \lambda_3 ) \kappa_1^2 \kappa_2^2  + \lambda_4 ( \kappa_1^3 \kappa_2  + \kappa_1 \kappa_2^3 ) +  \frac{1}{4} \rho_1 v_R^4 .
\end{align*}
Considering only the first stage of the transition, we assume $\kappa_1=\kappa_2=0$, giving:
\begin{equation}
    V_0 (v_R) = \frac{1}{2} \mu_3^2 v_R^2 + \frac{1}{4} \rho_1 v_R^4.
\end{equation}
We then replace all quantities with their 3d counterparts which have been matched during the dimensional reduction to obtain:
\begin{equation}
    V_{0,3}(v_{R,3}) = \frac{1}{2} \mu_{3,3}^2 v_{R,3}^2 + \frac{1}{4}\rho_{1,3} v_{R,3}^4.
    \label{eq::tree_3d}
\end{equation}
Note the mass term $\mu_3$ should be distinguished from $\mu_{3,3}$, where the subscript on the former does not indicate it is a 3d quantity.  Within \texttt{PhaseTracer}, we restore powers of temperature in the field and the potential to ensure the correct dimensionality.  This is done with the following matching conditions:
\begin{equation}
    V_3 = T V, \quad v_{R,3} = \frac{v_R}{\sqrt{T}}.
\end{equation}

The three-dimensional one-loop potential is derived within a Euclidean field theory framework, beginning from the summation over all one-loop 1PI diagrams to give:
\begin{equation}
    V_{1,3} = \frac{1}{2} \int \frac{ d^3 p }{ (2 \pi)^3 } \ln (p^2 + m^2).
\end{equation}
In the dimensional regularisation scheme, this becomes:
\begin{equation}
    V_{1,3} = - \frac{m^3}{2} \frac{1}{\frac{d}{2}( \frac{d}{2} - 1 )} \frac{\Gamma(2 - \frac{d}{2})}{(4 \pi )^{d/2}} \bigg (\frac{\mu^2}{m^2} \bigg )^{\frac{1}{2}(3 - d )}.
    \label{eq:OneLopeDimReg}
\end{equation}
This expression is finite for $d \rightarrow 3$, avoiding the pole in the $\Gamma$ function for $d=4$.  Restoring the correct dimensionality, we obtain the following closed-from expression for the one-loop potential:
\begin{equation}
    V_{1,3} = - \frac{m^3}{12 \pi}
\end{equation}
Which generalises for arbitrarily many scalar eigenstates to:
\begin{equation}
    V_{1,3} = - \frac{1}{12\pi} \sum_i n_i [m_{i, 3}^2]^{3/2},
\end{equation}
where $n_i$ accounts for the multiplicity of each eigenstate, as listed in the previous section.

\paragraph{Dimensional Reduction} \label{subsec::DimReg}

In constructing the effective field theory, we employed the Mathematica package \texttt{DRalgo}~\cite{Ekstedt:2022bff} to perform the dimensional reduction.  For brevity, we omit the full expressions of the matching relations for each parameter in the 3dEFT; instead, we summarise the structure of the relevant terms.  Particular care is taken to provide the explicit temperature dependence of the three-dimensional parameters.

Within the 3dEFT high-temperature expansion formalism all 3d parameters, e.g. $m^2_3$, $g^2_3$, $\lambda_3$, are functions of temperature, whilst the effective potential itself contains no explicit temperature dependence.  The matching relations that make explicit this temperature dependence have the following form:
\begin{alignat*}{3}
  & m_3^2 && = - m_0^2 + \alpha T^2 \quad && + \alpha' T^2, \\
  & \lambda_3 && = \lambda T && + \xi T, \\
  & g^2_3 && = g^2 T && + \zeta T .\\
  &  && \sim \mathcal{O}(\lambda) T && + \mathcal{O}(\lambda^2) T
\end{alignat*}
On the RHS all couplings belong to the parent 4d theory, and we have grouped these terms by their LO and NLO correspondence.  That is, when the matching is performed to leading order, 3d masses are evaluated at one-loop (with $\alpha \sim \mathcal{O}(\lambda)$), whilst couplings are taken at tree-level.  Note in the 3d theory, couplings gain a positive mass dimension, reflecting the super-renormalisable nature of the theory.  This LO approach is formally equivalent to utilising the high-temperature expansion in the conventional 4d approach.  At NLO, masses are evaluated at two-loop order, and couplings at one-loop.  The constants $\alpha, \alpha', \xi, \zeta$ are $\mathcal{O}(\lambda^2)$ functions of the 4d couplings $\lambda$ and $g^2$, the exact forms of which are particular to each theory.

Much like the effective potential, the one-loop expressions are UV-finite.  However, the two-loop 3d masses require renormalisation group methods.  The expressions for $\mu_{i,3}^2$ (that is, the 3d Lagrangian masses) will contain logarithms of the form:
\begin{equation}
    \ln \bigg ( \frac{Q_3}{Q} \bigg ),
\end{equation}
Here we employ the convention $Q = \pi T$, whilst we choose to match the 3d RG parameter $Q_3$ to the $U(1)_{B-L}$ Debye mass:
\begin{equation}
    Q_3 = \mu_{U(1)} = \sqrt{\frac{8 g_{BL}^2 T^2 }{3}}.
\end{equation}
This has the additional benefit of suppressing logs of the form $\ln(Q_3/\mu_G)$, where $G = SU(2)_L, \ SU(2)_R, \ U(1)$ are the vector Debye masses.  Such logarithms arise when integrating out temporal gauge bosons.

\section{Gravitational Waves} \label{sec::GravWaves}

In this work we employ the gravitational wave spectrum as calculated by \texttt{PhaseTracer}.  This produces a gravitational wave spectrum of the following form:
\begin{equation}
    \Omega h^2(f; \ \alpha, \beta/H, T_{\star}) = \Omega h^2_{\text{sw}} + \Omega h^2_{\text{turb}} + \Omega h^2_{\text{col}}.
\end{equation}
We consider gravitational waves from three sources: sound waves, magnetohydrodynamic (MHD) turbulence, and collisions.  Each spectrum is a function of the frequency, $f$, as well as four thermal parameters: $\alpha, \ \beta/H$, $T_{\star}$, and $v_w$. 

The bubble wall velocity, $v_w$, significantly influences the gravitational wave spectrum, affecting both its amplitude and spectral shape ~\cite{DeCurtis:2022hlx, DeCurtis:2023hil, DeCurtis:2024hvh, Branchina:2025jou}.  In this work, we treat $v_w$ as a fixed free parameter, with $v_w = 0.3$.  This simplification constitutes a notable limitation of our analysis.  However, the recent development of numerical tools such as \texttt{WallGo}~\cite{Ekstedt:2024fyq}, which enable automated computation of $v_w$, offer a promising avenue for future refinement.  We suspect that there is a large variation in $v_w$ across the parameter space, which may have a substantial impact on our results.

\subsection{Thermal Parameters}

\paragraph{Transition Strength and Energy Budgets}

The parameter $\alpha$, which quantifies the energy released during the phase transition available for gravitational wave production, serves as a measure of the transition's strength.  This is defined as~\cite{Athron:2023xlk}:
\begin{equation}
    \alpha = \frac{4(\theta_f(T_{\star}) - \theta_t(T_{\star}))}{3 w_f(T_{\star})},
    \label{eq::alpha_tr}
\end{equation}
where $\theta = \frac{1}{4}(\rho - 3 p)$ is the trace anomaly of each phase and $w_f$ the enthalpy density of the false vacuum.  The relevant hydrodynamic quantities can be derived from the partition function and are related to the effective potential $V$ by:
\begin{equation}
    p = - V, \quad \rho = T \frac{\partial p}{\partial T} - p = V - T \frac{\partial V}{\partial T}, \quad w = T \frac{\partial p}{\partial T} = - T \frac{\partial V}{\partial T}.
    \label{eq::hydro}
\end{equation}
We have normalised the one-loop effective potential such that $V = 0$ in the false vacuum.  As such, the only contribution to the pressure in the false vacuum comes from the radiation contribution, $p_R$, leading us to replace the enthalpy density with $w_f = \frac{4}{3} \rho_R$.  Using Eq.\eqref{eq::hydro}, the definition in Eq.\eqref{eq::alpha_tr} then becomes:
\begin{equation}
    \alpha = \frac{\Delta ( V - \frac{T}{4} \partial_T V )}{\pi^2 g_{\star} T^4/30},
\end{equation}
where $\Delta$ signifies the difference between the true $\phi_t$ and false $\phi_f$ vacuums, and $g_{\star}$ is the number of the relativistic degrees of freedom at $T_{\star}$.

The parameter $\alpha$ can be related to the kinetic energy fraction $K$, by defining the efficiency coefficient $\kappa$ such that~\cite{Kamionkowski:1993fg}:
\begin{equation}
    K = \frac{\kappa \alpha}{1 + \alpha}.
\end{equation}
This relates the energy released by the phase transition $\alpha$ to the fluid kinetic energy available to generate gravitational radiation.  The definition of $\kappa$ for each source of gravitational waves can be found in Appendix \ref{App::GW}.

\paragraph{Bounce Action}

The decay rate of the metastable false vacuum is given by~\cite{Coleman:1977py, Callan:1977pt, Linde:1981zj}:
\begin{equation}
    \Gamma = A(T) e^{-B(T)},
\end{equation}
where the values for both $A, B$ depend on the nature of the decay.  For a decay via thermal fluctuations, $B(T) = S_3(T)/T$, where $S_3(T)$ is the three-dimensional Euclidean bounce action.  This is evaluated within \texttt{PhaseTracer} by employing the path deformation method~\cite{Wainwright:2011kj}.  The above expression for the decay rate employs a semi-classical approximation to a full path integral expression, and as such the prefactor $A(T)$ stems from fluctuations about the classical action, and can be expressed in the form:
\begin{equation}
    A(T) = \left ( \frac{B(T)}{2\pi} \right )^{3/2} D(T),
\end{equation}
where $B(T)$ is as above.  The functional determinant, $D(T)$, is approximated using $D(T) \sim T^4$.  This approximation is expected to introduce leading-order uncertainty into our calculation of the gravitational wave spectrum.  The existence of numerical codes for the evaluation of this determinant, such as \texttt{BubbleDet}~\cite{Ekstedt:2023sqc}, provide a key source of improvement.  We thus obtain for the decay rate:
\begin{equation}
    \Gamma(T) = T^4 \left ( \frac{S_3(T)}{2\pi T} \right )^{3/2} e^{-S_3(T)/T}. \label{eq::DecayRate}
\end{equation}
This full expression will be relevant to the study of the false vacuum fraction below.  For estimating gravitational waves, we linearise the bounce action $S_3$ about the transition temperature $T_{\star}$ (see Eq.\eqref{sec::Tref}) to obtain:
\begin{equation}
    \Gamma(T) \approx \Gamma(T_{\star}) e^{- \beta(T_{\star}) (T - T_{\star})},
\end{equation}
where $\beta(T_{\star})$ is given by:
\begin{equation}
    \beta(T) = T H(T) \frac{d}{dT} \bigg ( \frac{S_3(T)}{T} \bigg ) \Rightarrow \frac{\beta(T_{\star})}{H(T_{\star})} = T \frac{d}{dT} \bigg ( \frac{S_3(T)}{T} \bigg ) \bigg |_{T = T_{\star}}.
    \label{eq::betaH}
\end{equation}
The quantity $\beta/H$ appears in the gravitational wave evaluations.

\paragraph{Transition Temperature} \label{sec::Tref}

The progress of our phase transition is contained within the false vacuum fraction, $P_f(T)$, giving the fraction of the universe still remaining in the false vacuum.  Thus, we expect $P_f(T_c) = 0$, whilst $P_f(T) \rightarrow 1$ for $T \ll T_c$.  The false vacuum fraction is given by~\cite{Athron:2022mmm}:
\begin{equation}
    P_f(T) = \exp \left ( - \frac{4 \pi }{3} v_w^3 \int_T^{T_C} dT' \frac{\Gamma(T')}{T'^4 H(T')} \left (\int_T^{T'} dT'' \frac{1}{H(T'')}  \right)^3 \right ).
\end{equation}
We can simplify this expression under the assumption of a sufficiently fast transition (that is, no supercooling).  In this case, we assume the contribution to the Friedmann equation from the vacuum energy is negligible, leaving only the dominant term from the radiation density.  This assumption is founded by previous investigations of the LRSM, which show that the model favours fast transitions with small $\alpha$ and large $\beta/H$.  Thus, the Friedmann equation reads:
\begin{equation}
    H^2 = \frac{8 \pi G}{3} \rho_R = \frac{8 \pi^3 g_{\star}G}{90} T^4 = C^2 T^4,
\end{equation}
where we have defined $C = \sqrt{8\pi^3 g_{\star}G/90}$ and the number of the relativistic degrees of freedom is $g_{\star} \sim 134$~\cite{Brdar:2019fur}.  This allows the inner integral to be evaluated analytically to give:
\begin{equation}
    P_f(T) = \exp \left ( - \frac{4 \pi }{3 C^4} v_w^3 \int_T^{T_C} dT' \frac{\Gamma(T')}{T'^6} \left (\frac{1}{T} -  \frac{1}{T'} \right)^3 \right ) ,
\end{equation}
removing the nested integration over $H(T)$. 

We define three temperature milestones in this work.  The first is the nucleation temperature, $T_n$, defined as the temperature at which one bubble is nucleated per Hubble volume of order $H^{-3}$.  This condition is expressed in terms of the nucleation rate, $N(T)$, as:
\begin{equation}
    N(T_n) = \frac{4 \pi}{3} \int_{T_n}^{T_c} dT' \frac{\Gamma(T') P_f(T')}{H^3(T')} = 1.
\end{equation}
As above, we have $H(T) = C T^2$.  We assume $P_f(T)$ does not deviate meaningfully from 1 over the integration domain above.  For transitions without significant supercooling, $P_f(T)$ rapidly drops from $1$ to $0$ over a short temperature range.  Provided $T_n > T_p$ this assumption is valid.  The nucleation condition then simplifies to:
\begin{equation}
    \frac{4 \pi}{3} \int_{T_c}^{T_n} dT' \frac{\Gamma(T')}{H^3(T')} = 1.
    \label{eq::NucTemp}
\end{equation}
As with the assumption on $H(T)$, this approximation removes a computationally expensive nested integral at the cost of compromising the accuracy of $T_n$.  The integration can be removed entirely by defining the nucleation temperature as the point where one bubble is nucleated in a Hubble volume of order $H^{-3}$ in a given Hubble time $H^{-1}$, given by:
\begin{equation}
    \frac{\Gamma(T_n)}{H^4(T_n)} = 1.
\end{equation}
A more common heuristic for approximating the nucleation temperature is defined in terms of the bounce action in the constraint:
\begin{equation}
    \frac{S_3(T_n)}{T_n} \approx 140.
\end{equation}
This equation assumes that the transition occurs during the radiation dominant epoch and at the electroweak scale.  The former assumption is valid, however, the latter is violated for the LRSM where $v_R \gg v_h$.  This is likely a significant source of error in the gravitational wave estimate, and as such, we utilise the assumption $P_f(T) \sim 1 $ for $T_c > T > T_n$ and employ ~Eq.\eqref{eq::NucTemp}.

The above assumption for the nucleation temperature means that it is formally decoupled from the progress of the transition, which is best captured in the evolution of $P_f(T)$.  This decoupling is of primary concern for transitions with significant supercooling, where sparsely nucleated but fast-growing bubbles can cause $P_f(T)$ to deviate significantly from unity during the evaluation of $N(T)$.  To address this issue, we also employ the percolation temperature, $T_p$.  For the case of finite volume (computationally this is the finite of the simulation domain, physically it corresponds to the Hubble volume), percolation is defined as the point wherein a complete cluster of bubbles spans the medium.  That is, there exists a path across the medium remaining entirely within nucleated bubbles.  Simulations show this condition is in correspondence with the temperature:
\begin{equation}
    P_f(T_p) \approx 0.71.
    \label{eq::PercTemp}
\end{equation}

Lastly, we check that the transition completes by solving for the completion temperature $P_f(T_f) \rightarrow 0$.  The exponential nature of the false vacuum fraction means that such a condition is never reached for finite temperature.  As such, we impose the completion condition:
\begin{equation}
    P_f(T_f) < \epsilon \label{eq::CompTemp},
\end{equation}
and take $\epsilon = 0.01$.

We find that in all cases unit nucleation (given by Eq.\eqref{eq::NucTemp}) is achieved, both percolation (given by Eq.\eqref{eq::PercTemp}) and completion (given by Eq.\eqref{eq::CompTemp}) are also achieved, and vice versa.  This behaviour is expected for a weak and fast transition, which we observe from the calculated values of $\alpha$ and $\beta/H$.  Additionally, we find little variation between the nucleation and percolation temperatures.  As such, all gravitational wave results below have been given using $T_\star = T_p$.

\subsection{Fitting Formulas}
\label{sec::fitting}

Determining the gravitational wave spectra from a cosmological phase transition is an ongoing and open problem in particle cosmology.  The most accurate approaches require complicated hydrodynamical simulations of the scalar-plasma system.  Such simulations are well beyond of the scope of this study.  Instead, we employ the approximations provided by $\texttt{PhaseTracer}$ in the form of fitting formulas matched to these simulations.  A full account of these formulas can be found in~\cite{Athron:2024xrh} and references therein.  A brief account of these formulas are given below, and the exact expressions are included for reference in Appendix \ref{App::GW}.

Gravitational waves sourced from bubble wall collisions are found using the envelope approximation.  The resulting spectrum for both collisions and sound walls has the following general form:
\begin{equation}
    \Omega_i h^2 = \mathcal{A}_i(v_w) \left ( \frac{\beta_{\star}}{H_{\star}} \right )^{-n_{\beta, i}} \left ( \frac{\kappa_i \alpha }{1 + \alpha} \right )^{n_{\alpha, i}} \left ( \frac{g_\star}{100} \right )^{-1/3} \mathcal{S}_i(f/f_{\text{peak},i}).
\end{equation}
Here, the index $i$ runs over collisions and turbulence.  The amplitude $\mathcal{A}$, indices $n_\beta$ and $n_\alpha$, the efficiency factor $\kappa$, peak frequency $f_{\text{peak}}$, and spectral functions $\mathcal{S}$ can all be found in Appendix \ref{App::GW}.

Sound waves are assumed to be the dominant source of gravitational wave radiation.  The resultant spectrum has the same form as above with an additional suppression term in the amplitude factor, given by:
\begin{equation}
    \begin{split}
    \Omega_{\text{sw}} h^2 = & \mathcal{A}_{\text{sw}}(v_w) \left ( \frac{\beta_{\star}}{H_{\star}} \right )^{-n_{\beta, {\text{sw}}}}\left ( \frac{\kappa_{\text{sw}} \alpha }{1 + \alpha} \right )^{n_{\alpha, {\text{sw}}}} \left ( \frac{g_\star}{100} \right )^{-1/3} \\
    & \times \mathcal{S}_{\text{sw}}(f/f_{\text{peak},{\text{sw}}}) \times \tl{{\Omega}}_{\text{gw}} \min(H_\star R_\star/\overline{U}_f, 1).
    \end{split}
\end{equation}
The additional $\min(H_\star R_\star/\overline{U}_f, 1)$ term accounts for shock formation.  Note our presentation of this formula differs from that provided in~\cite{Athron:2024xrh}.  A derivation of their equivalence is given in Appendix \ref{App::GW}.

\subsection{Signal to Noise Ratio}
\label{sec::snr}

In this work, we assess detectability of the resultant gravitational signals using two metrics: comparison against the \textit{peak-integrated sensitivity curve} (PISC), and the \textit{signal to noise ratio} (SNR).  Utilisation of the PISC will serve as the primary metric, and can be understood as follows.  Given a gravitational wave spectrum $\Omega h^2$, the peak amplitude and frequency are projected into the $(\Omega h^2_{\rm peak}, f_{\rm peak})$ plane.  These points are then compared against the PISC.  Those lying above the curve suggest a sufficiently large SNR to be detected.  In particular, the PISC curve is defined as the envelope of all peak points for spectra $\Omega h^2$ with a SNR of $1$~\cite{Schmitz:2020syl}. 

We use the approximations to the PISC for BBO and DECIGO provided in Section 7.3 of~\cite{Schmitz:2020syl}.  It should be noted that these are semi-analytical fitting functions and depend on the specific shape of the gravitational wave spectrum $\mathcal{S}_i$, which accounts for inconsistencies between the relative positions of our PISC and our quoted SNR, the latter being calculated with our own $\mathcal{S}_i$.  To determine the SNR, $\rho$, we adopt the conventional prescription and define:
\begin{equation}
    \rho^2 = 2 T_{obs} \int df \left ( \frac{\Omega_{\rm gw}(f)}{\Omega_{\rm sens}(f)} \right)^2 ,
\end{equation}
where $\Omega_{\rm gw}$ is the sum over each amplitude stated above.  The noise energy density power spectrum is related to the effective strain noise power spectral density $S_{\rm eff}(f)$ as:
\begin{equation}
    \Omega_{\rm sens}(f) = \frac{2 \pi^2}{3 H_0^2} f^3 S_{\rm eff}(f).
\end{equation}
For BBO, $S_{\rm eff}$ can be related to the detector power spectral density $P_n(f)$ and the response functions $R (f)$.  $P_n(f)$ is defined as:
\begin{equation}
    P_n(f) = \frac{4}{L^2} \bigg [ (\delta x)^2 + \frac{(\delta)^2}{(2 \pi f)^4} \bigg ],
\end{equation}
where $L = 5 \times 10^7$m is the detector arm length, while $(\delta x)^2 = 2 \times 10^{-34}$ m$^{2}$Hz$^{-1}$ and $(\delta a)^2 = 9 \times 10^{-34}$ m$^{2}$s$^{-4}$Hz$^{-1}$ are the position and acceleration noise respectively~\cite{Crowder:2005nr}.  Lastly, the response function $R(f)$ is related to the normalised overlap reduction function $\gamma(f)$ by:
\begin{equation}
    R(f) = \frac{\sin^2(\delta)}{5} \gamma(f),
\end{equation}
where $\delta = \pi/3$ is the detector arm angle.  The overlap reduction function has been obtained from the numerical results provided in~\cite{Thrane:2013oya}.  We assume an observation time of $T_{obs} = 3$ years.

\section{Methodology}
\label{sec::analysis}

The LRSM as outlined in Section~\ref{sec::Model} has the following 22 parameters:
\begin{equation}
    \{ g_L, \ g_{BL}, \ y_1, \ y_2, \ \mu_{1,2,3}, \ \alpha_{1,2,2^*,3}, \ \beta_{1,2,3}, \ \rho_{1,2,3,4}, \ \lambda_{1,2,3,4} \},
\end{equation}
where we have made the assumptions $g_L = g_R$, $\mathcal{F}_{ij} = \text{diag}(0, 0, y_1)$ and $\mathcal{G}_{ij} = \text{diag}(0, 0, y_2)$ as motivated in the previous section.  In Appendix \ref{App::VEV} we show that our chosen vacuum alignment requires the simplifying assumptions $\alpha_2 = \alpha_2^*$ and $\beta_i = 0$, which further reduces the number of free parameters.  The scalar masses $\mu$ are fixed in terms of the VEVs $\kappa_+$, $r$, and $v_R$.  The first of these is matched to the electroweak VEV $v_h = 246$ GeV.  The latter two, $r$ and $v_R$, are free parameters.  In deriving our scalar mass eigenstates in Appendix \ref{App::Eigens}, we have assumed $r$ to be perturbatively small and as such we will assume $r = 10^{-3}$ in this work.  The $\mathcal{P}$-breaking VEV $v_R$ is much more constrained.  Most masses in this model (with few exceptions) scale with $v_R$, as such, if this VEV is too low it leads to BSM physical states present at the electroweak scale.  The easiest lower bound on $v_R$ is derived from the aforementioned left-handed neutrino mass.  To ensure a sub-eV mass, we impose $v_R \geq 10^4$ GeV.  Detection of this transition at near future observatories will favour a low $v_R$ scale, as pushing $v_R$ higher drives the characteristic frequency beyond the few Hz range.  Thus, we set $v_R = 10$ TeV.

From the discussion in the previous section, the parameters $\{ g_L, g_{BL}, y_1, y_2, \lambda_1 \}$ can all be matched to SM counterparts.  The $SU(2)_L$ gauge coupling is unchanged during the $\mathcal{P}$ transition, and is matched to the value of $g_L$ after running to $g_L(Q = v_R)$.  Likewise, $g_{BL}$ can be matched to $g_Y$ utilising the following result~\cite{Peskin:1995ev}:
\begin{equation}
    g_Y = \frac{g_R g_{BL}}{\sqrt{g_R^2 + g_{BL}^2}},
\end{equation}
where $g_R=g_L$.  The above expression is exactly analogous to the SM EWSB case.  For $r\ll 1$, the Yukawa couplings $y_1$ and $y_2$ can be matched to their SM counterparts $y_t$ and $y_b$.  Lastly, given any values of $\alpha_1$ and $\rho_1$ we constrain $\lambda_1$ to $\lambda_{SM}$ using Eq.\eqref{eq::lambda1constraint}.

Thus, we first take the SM values of $\{g_L, g_Y, y_t, y_b, \lambda_1\}$, which we take to be defined at the scale $Q = m_Z$ to be: $\{0.6534, 0.3503, 0.9932, 0.0240, 0.1290\}$.  These are then run to the $\mathcal{P}$-breaking scale $Q = v_R$ using the one-loop SM beta functions (which we obtained using \texttt{DRalgo}) and matched to the LR parameters as outlined above.  This leaves the following 10 free-parameters to be defined at $Q = v_R$:
\begin{equation}
    \{ \alpha_{1,2,3}, \ \rho_{1,2,3,4} , \ \lambda_{2,3,4} \}.
\end{equation}

This parameter space is then restricted by a set of physicality conditions.  Firstly, the chosen parameters must correspond to a `good' vacuum of the LRSM ~\cite{BhupalDev:2018xya}.  The VEV alignment described in Eq.\eqref{eq:VEValign} partly ensures this, however, the additional limits have been shown to enhance the probability of a good vacuum structure~\cite{BhupalDev:2018xya}:
\begin{align}
    & 0 \leq \mu_i^2 \leq 4 \pi v_h^2 && i \in \{1, 2, 3 \}, \\
    & 0 \leq \lambda_i, \ \rho_i \leq 4 \pi && i \in \{1, 2, 3, 4 \}, \\
    & 0 \leq \alpha_i \leq 0.2 \pi && i \in \{1, 2, 3 \}, \\
    & \beta_i = 0 && i \in \{1, 2, 3 \}.
\end{align}
Note that the requirement $\beta_i=0$ has been enforced irrespective of these constraints (see Appendix \ref{App::VEV}).  In addition, we impose the lower bound $\lambda' > 0$ for all quartic couplings $\lambda'$\footnote{In this section, $\lambda'$ is used to represent an arbitrary quartic coupling of our model.  It should be distinguished from the Higgs self-interactions $\lambda_{1,2,3,4}$.}.  These restrictions are largely made redundant in this analysis, as we have used symmetry-breaking conditions and a restricted vacuum structure (e.g. $v_L=0$).  Failure to land in a valid vacuum is identified by checking for a failure in the phase finding methods utilised in \texttt{PhaseTracer}, and in such cases the parameter point is disregarded.

The tree-level potential with our chosen VEV alignment is a quartic polynomial in three variables $\kappa_1$, $\kappa_2$, and $v_R$.  The following constraints ensure this potential is bounded from below~\cite{Chakrabortty:2013mha, Chakrabortty:2013zja}:
\begin{equation}
    \lambda_1 \geq 0, \quad \rho_1 \geq 0, \quad \rho_1 + \rho_2 \geq 0, \quad \rho_1 + 2 \rho_2 \geq 0.
\end{equation}

Given the mass eigenstates in Appendix \ref{App::Eigens}, positive-definite physical Higgs masses can be imposed by maintaining:
\begin{equation}
    \rho_3 - 2 \rho_1 > 0 .
\end{equation}
Otherwise, positive-definite masses are ensured by considering only positive Higgs couplings $\lambda' \geq 0$.

We impose unitary bounds derived from the contact $H_i H_j \rightarrow H_k H_l$ scattering process, where $H_i$ represents any of the physical scalar fields.  Writing the scalar potential in our physical basis:
\begin{equation}
    V = \sum_{\{i,j,k,l \} } \Lambda_m H_i H_j H_k H_l,
\end{equation}
unitary bounds can be satisfied by ensuring $|\Lambda_m| < 8 \pi$.  This establishes the following limits~\cite{Chakrabortty:2016wkl}:
\begin{equation}
    \begin{split}
        \lambda_1 & < 4 \pi /3, \quad (\lambda_1 + 4 \lambda_2 + \lambda_3) < 4 \pi, \\
        (\lambda_1 & - 4 \lambda_2 + 2 \lambda_3) < 4 \pi, \\
        \lambda_4 & < 4 \pi / 3 \\
        \alpha_1 & < 8 \pi, \quad \alpha_2 < 4 \pi, \quad (\alpha_1 + \alpha_3) < 8 \pi, \\ 
        \rho_1 & < 4 \pi / 3, \quad (\rho_1 + \rho_2) < 2 \pi, \quad \rho_2 < 2 \sqrt{2} \pi , \\
        \rho_3 & < 8 \pi , \quad \rho_4 < 2 \sqrt{2} \pi .
    \end{split}
\end{equation}

In constructing the 3dEFT, we perform the matching at scales $Q = \pi T$, for $T \sim v_R$.  As such, the LRSM parameters are run from $Q = v_R$ to $Q = 10 v_R$.  For large initial conditions at $Q \sim v_R$ (e.g. $\lambda' \sim \pi$) we find our parameter points rapidly exceed the unitary and perturbative limits, or worse, begin to diverge as we encounter a Landau pole~\cite{Chauhan:2018uuy, Rothstein:1990qx}.  When this is encountered, the parameter point is disregarded.  This constraint proves to be the most restrictive for generating physically meaningful gravitational wave predictions.  We find the strongest gravitational wave amplitudes correspond to large values of $\alpha_1$ and $\rho_{1,3}$, which correspondingly have the strongest tendency to diverge.  A failure to account for this constraint would thus lead to overestimating the maximum gravitational wave predictions allowed within this model, potentially by several orders of magnitude.

\subsection{The MLS Method}
\label{sec::MLS}

The constrained ten-dimensional parameter space above poses a challenge for investigating the gravitational wave landscape of this phase transition.  The curse of dimensionality is compounded by the $\mathcal{O}(1)$ second evaluation times required to fully calculate the gravitational wave spectrum for each parameter point.  To overcome this issue, we adopt a variation of the \textit{Machine Learning Scan} (MLS) method originally proposed in~\cite{Ren:2017ymm}.  Our procedure is as follows.

To begin, we generate an initial data set, $\mathcal{D}_0$, by random sampling of the constrained parameter space.  From the onset, we run into the issue of obtaining a high enough resolution across the ten available dimensions.  For example, to reasonably sample observables over these ten dimensions, a simple discretisation of ten samples per axis would result in an approximately $10^{10}$ second evaluation time, or over 300 CPU years.  We have no means of circumventing this issue without adopting an improved sampling method.

Thus, from this initial data set, we proceed to the first iteration and construct a training set $\Gamma_1 = \mathcal{D}_0$, containing the quartic parameters as input, and the (logarithmic) peak gravitational wave amplitude as output.  A deep neural network $\mathcal{M}_1$ is then trained on this data and used to determine which part of the parameter space leads to the strongest gravitational wave amplitudes.  From this information, we generate a set of suggested predictions $\mathcal{P}_1$.  These suggested parameter points are combined with a set of new randomly generated samples before being passed to the actual \texttt{PhaseTracer} gravitational wave calculator to generate $\mathcal{D}_1$, the \textit{true} gravitational wave results. 
Random samples are added to force broader exploration, avoiding over-focusing on known regions.  This process is then repeated iteratively using the new training data $\Gamma_2 = \mathcal{D}_1 \cup \mathcal{D}_0$, or $\Gamma_j = \mathcal{D}_{j-1} \cup \mathcal{D}_{j-2}$.

The neural network architecture we use contains six hidden layers which begin with a 64-neuron layer, followed by a sequence of four 128-neuron layers before a final 64-neuron layer.  Each hidden layer uses the ReLU activation function and applies L2 regularisation with a factor of 0.001.  Additionally, a dropout layer with a rate of 0.3 follows each layer.  The final layer consists of a single neuron with linear activation.  We train the network without validation to 5000 epochs using the Adam optimiser with mean squared error loss.  No early stopping condition is employed.

\section{Results}
\label{sec::results}

\begin{figure}[htbp]
    \centering
    \includegraphics[width=\linewidth]{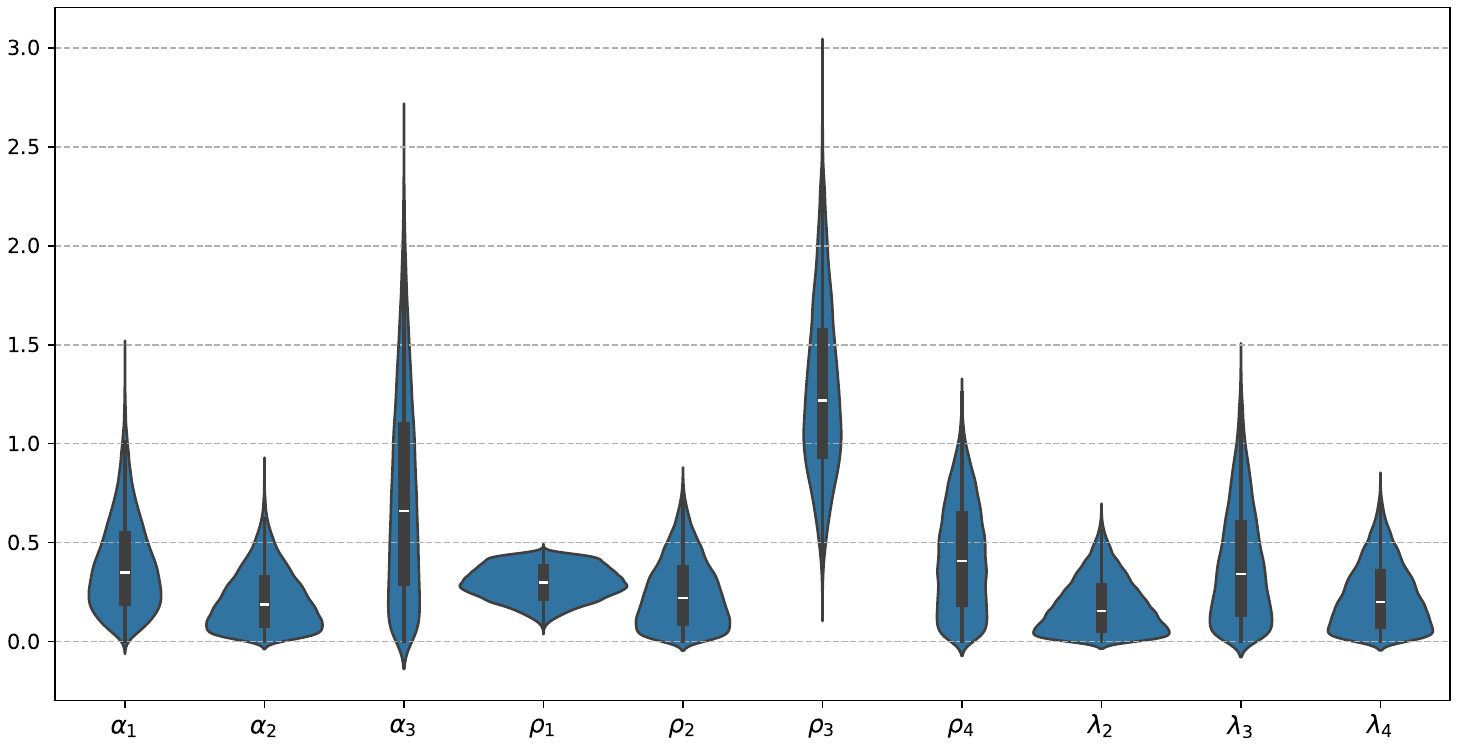}
    \caption{Violin plot showing the distribution of the investigated parameter space of the preliminary dataset $\mathcal{D}_1$, consisting of 19,813 samples.}
    \label{fig:initial_violin}
\end{figure}

The parameter space identified for the initial training data is shown in Fig.\ref{fig:initial_violin}.  These regions were found by randomly sampling parameters in the domain $[0, 2 \pi]$ and testing the results against previously discussed Higgs constraints.  We find that some parameters, such as $\alpha_3$ and $\rho_3$, allow for a wide available range, extending as far as $\lambda \approx \pi$.  However, in general, most parameters are limited to the domain $[0, 0.5]$.  It should be noted that the sampling efficiency of this approach is extremely limited.  Due to the heavy constraints on the parameter space, most randomly drawn samples fail to pass all of the checks.

\begin{figure}[htbp]
    \centering
    \includegraphics[width=0.6\linewidth]{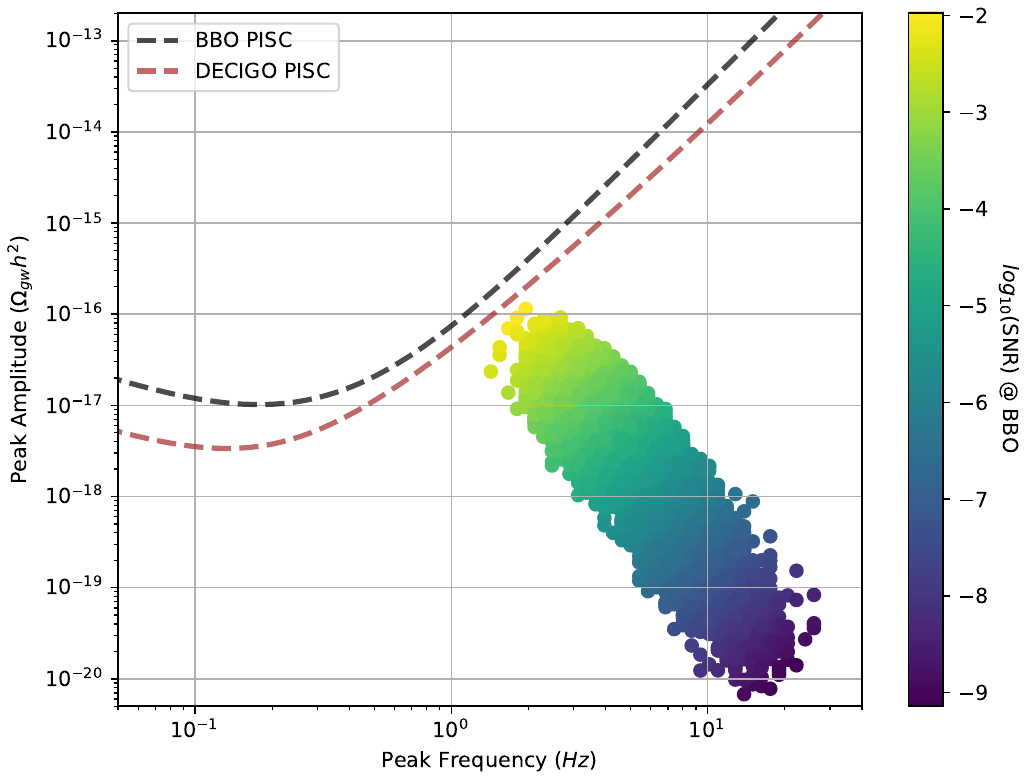}
    \caption{Gravitational wave estimates across the initial parameter space. Also shown are the PISC for BBO and DECIGO. These results show minimal prospects for detectability at these observatories within the initial parameter set.}
    \label{fig:initial_data}
\end{figure}

In Fig.\ref{fig:initial_data}, the gravitational wave predictions are calculated across the initial parameter space.  Shown are the peak amplitude and peak frequency of the full gravitational wave spectrum.  In addition, we provide the SNR at BBO as well as the PISC at both BBO and DECIGO~\cite{Schmitz:2020syl} to assess the detection prospects of these results.  Comparing these results to previous investigations of the triplet LRSM~\cite{Brdar:2019fur}, we find our predictions for the peak amplitude to be generally much lower.

These differences are likely attributed to two causes: tighter constraints on the overall parameter space and the state-of-the-art dimensional reduction approach.  At one loop, a strong gravitational wave signal $(\alpha \gg 1)$ requires contributions from either or both of the gauge boson and/or scalar eigenstates.  For the former to hold, we require a large separation of scales between the gauge $g$ and scalar $\lambda$ couplings.  Such a hierarchy is forbidden within this analysis, as the dimensional reduction requires $g^2 \sim \lambda$.  For the latter, we predict that parts of the parameter space for which the scalar eigenstates generate a large enough first-order barrier are forbidden by our constraints.

\begin{figure}[htbp]
    \centering
    \begin{subfigure}[b]{\textwidth}
        \centering
        \includegraphics[width=\textwidth]{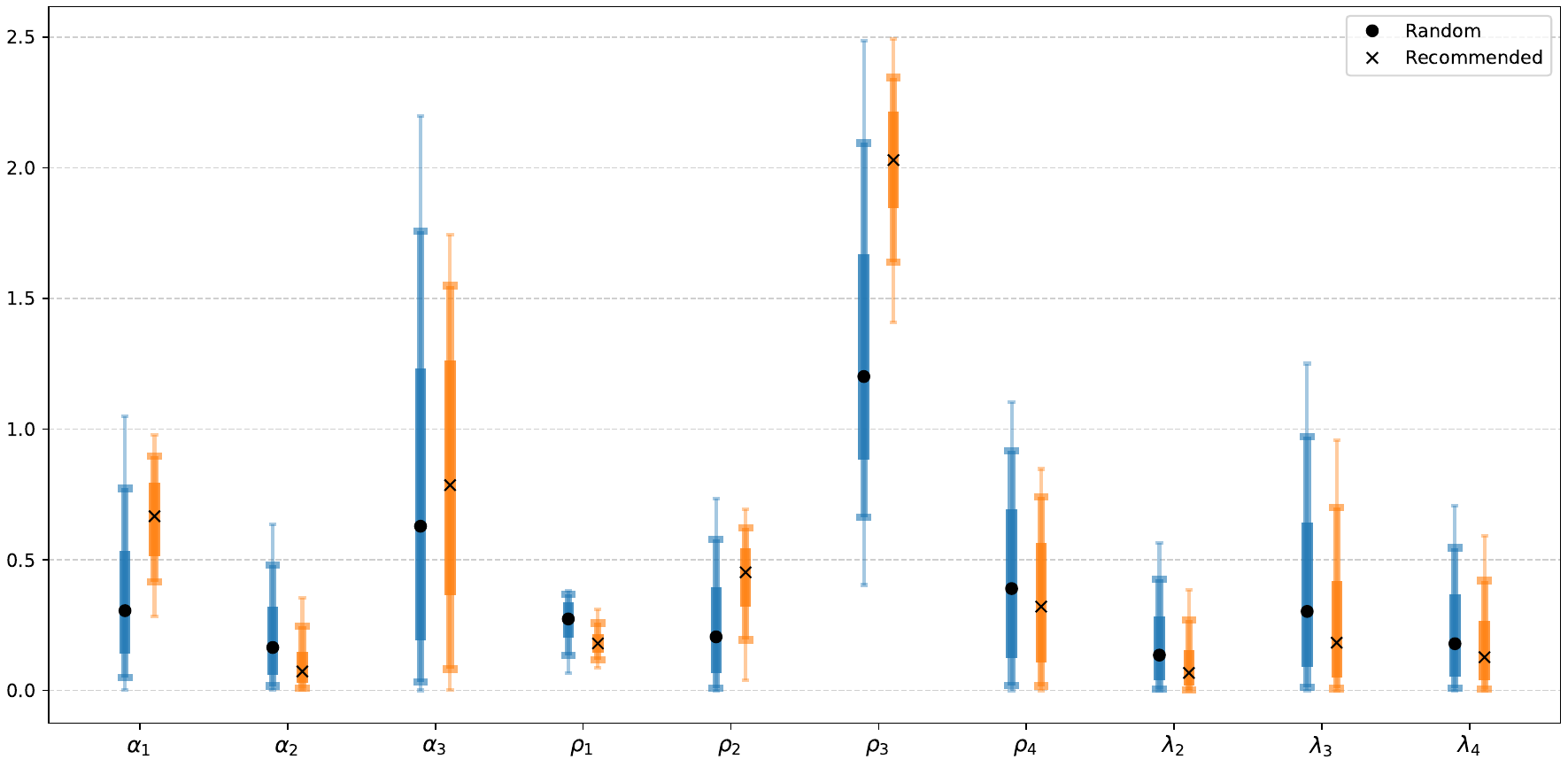}
    \end{subfigure}
    \vfill
    \begin{subfigure}[b]{\textwidth}
        \centering
        \includegraphics[width=\textwidth]{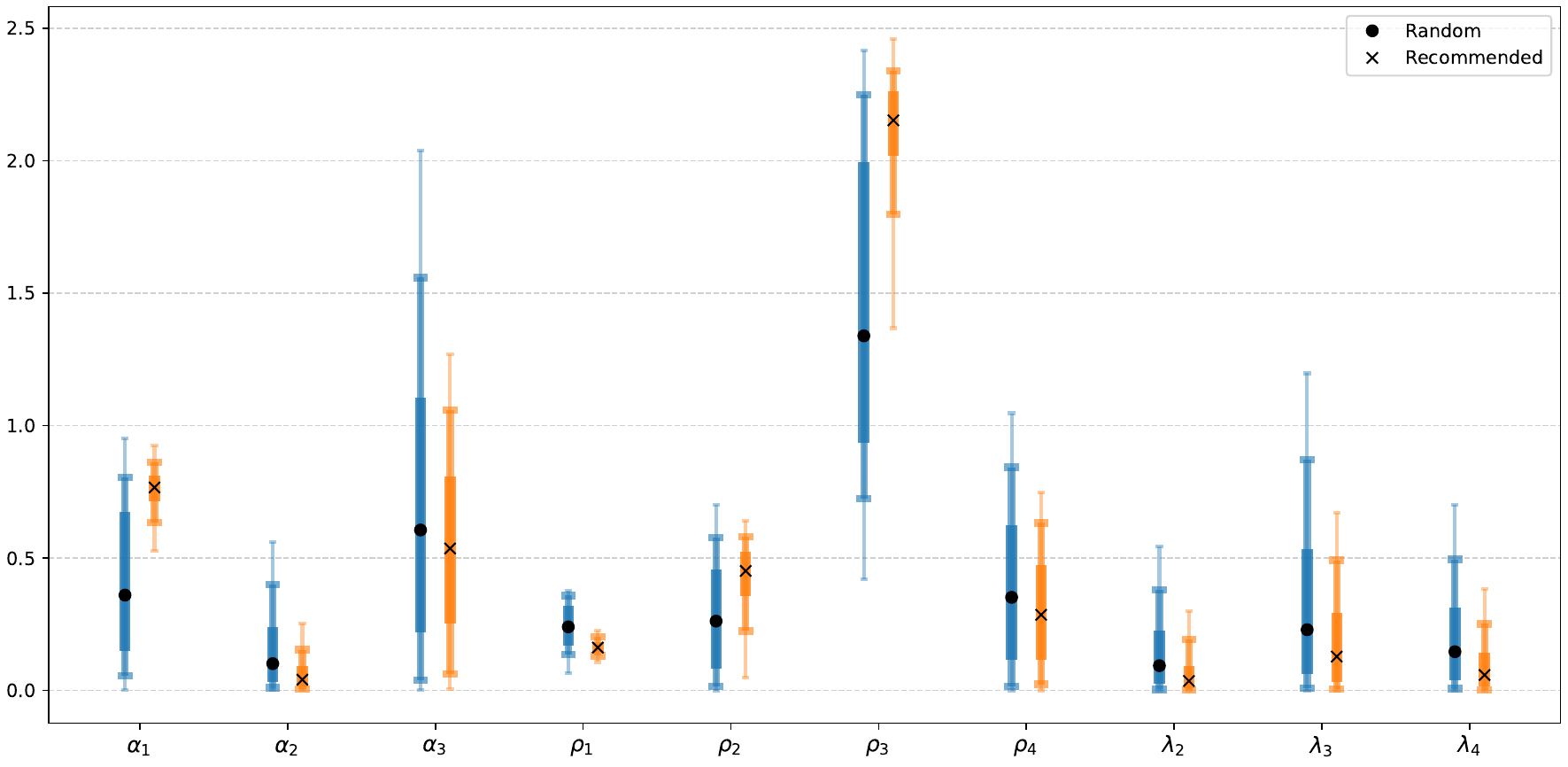}
    \end{subfigure}
    \caption{Comparison of random sampling (blue) against the recommended parameter points (orange) for the first and final iterations.}
    \label{fig:bar_comp}
\end{figure}

The stricter results presented in Fig.\ref{fig:initial_data} appear to rule out the detectability of the left-right phase transition. However, we now employ the MLS sampling scheme to demonstrate that this is not the case.  Proceeding with the method explained in Section \ref{sec::MLS}, we consider five iterations to determine if the preliminary results established above can be improved upon.  Fig.\ref{fig:bar_comp} compares the random parameter space to the recommended parameter space (that is, the set of points recommended by the NN) across the initial and final iterations.  In general, there is agreement between the two sets.  However, for parameters $\alpha_1$ and $\rho_3$, and to a lesser extent $\rho_2$, the domain of the two parameter sets begins to show a significant deviation.  The bulk of the parameter recommendations lay within only an outer quartile of the random parameter space.  This demonstrates the marked inefficiency of the random sampling scheme in accurately detecting the regions where the strongest gravitational signal exists.  Conversely, it demonstrates the efficiency of the NN in identifying regions where the strongest gravitational wave signals can be found.


\begin{figure}[htbp]
    \centering
    \includegraphics[width=\linewidth]{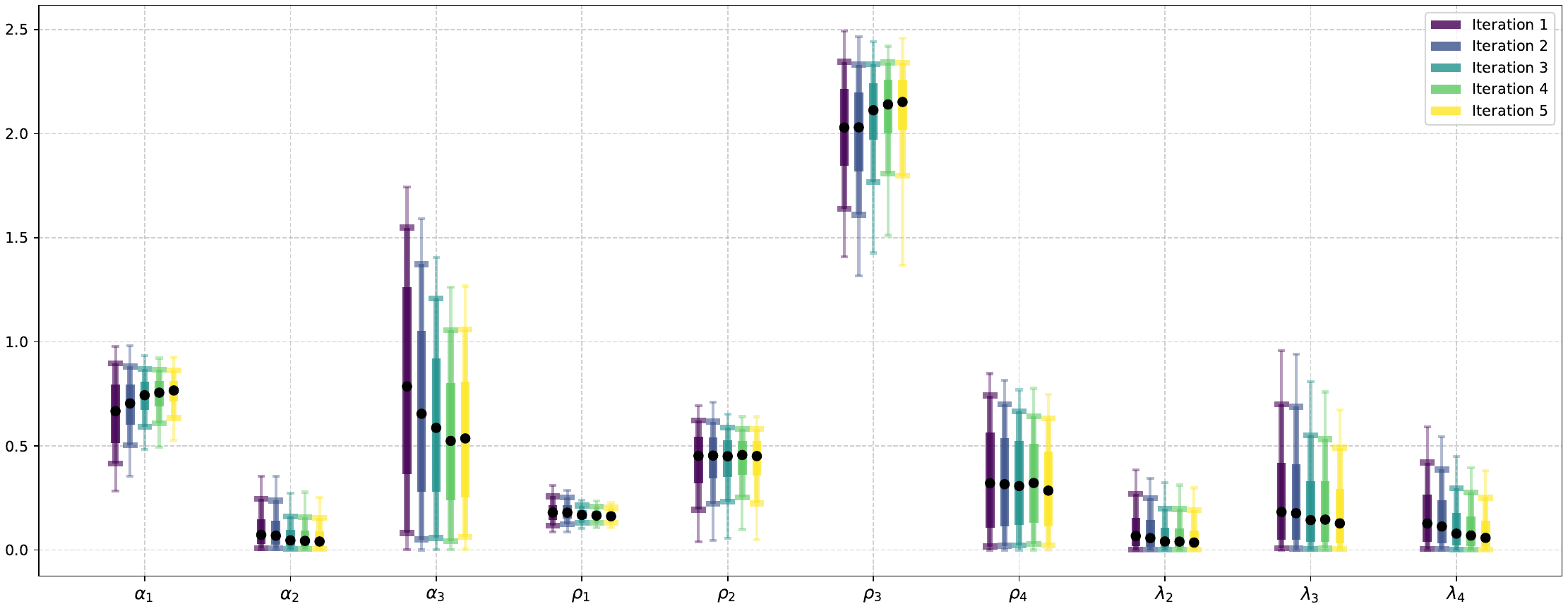}
    \caption{Comparison of parameter recommendations across each iteration. The NN can be seen to converge to a consistent domain for most parameter recommendations by the third iteration. Convergence is still seen in the lambda-type couplings by the final iteration, indicating headroom for further improvement on gravitational wave estimates.}
    \label{fig:iteration_comp}
\end{figure}

\begin{figure}[htbp]
    \centering
    \begin{subfigure}[b]{0.49\textwidth}
        \centering
        \includegraphics[width=\textwidth]{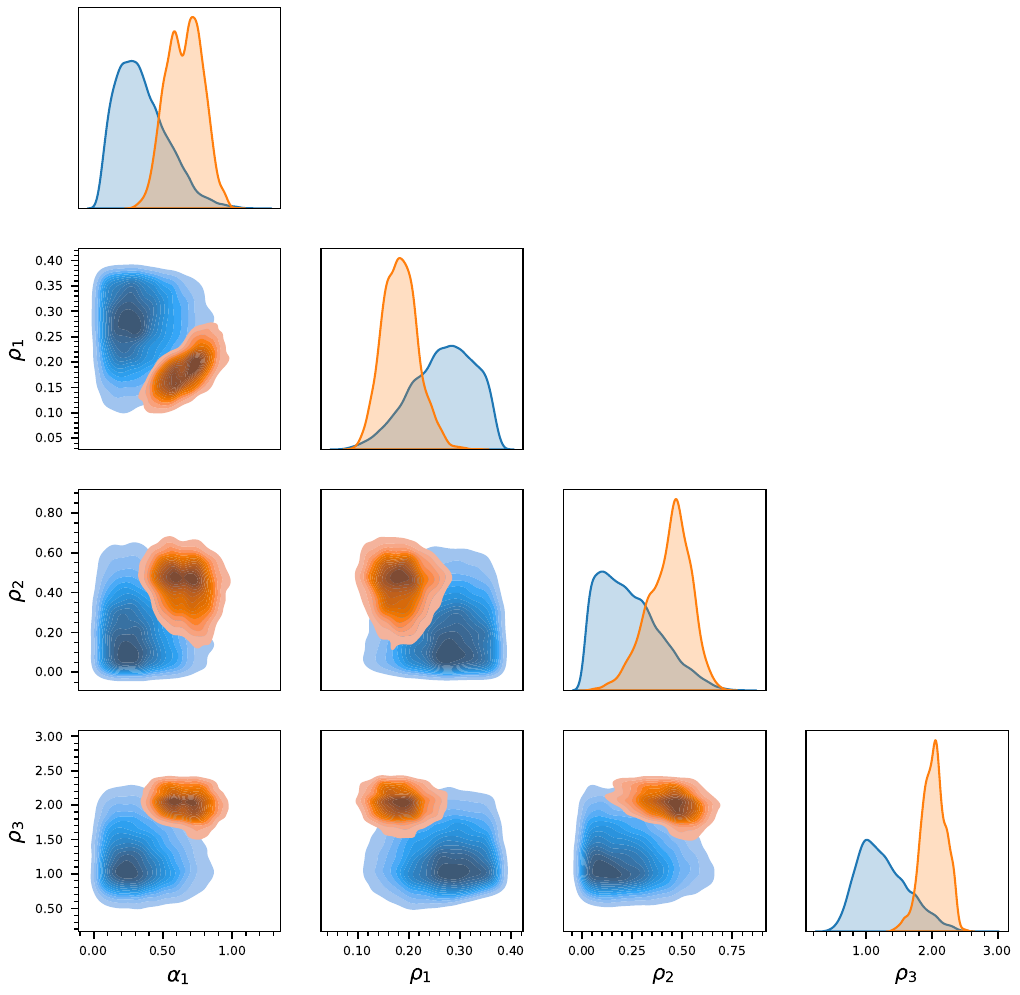}
    \end{subfigure}
    \hfill
    \begin{subfigure}[b]{0.49\textwidth}
        \centering
        \includegraphics[width=\textwidth]{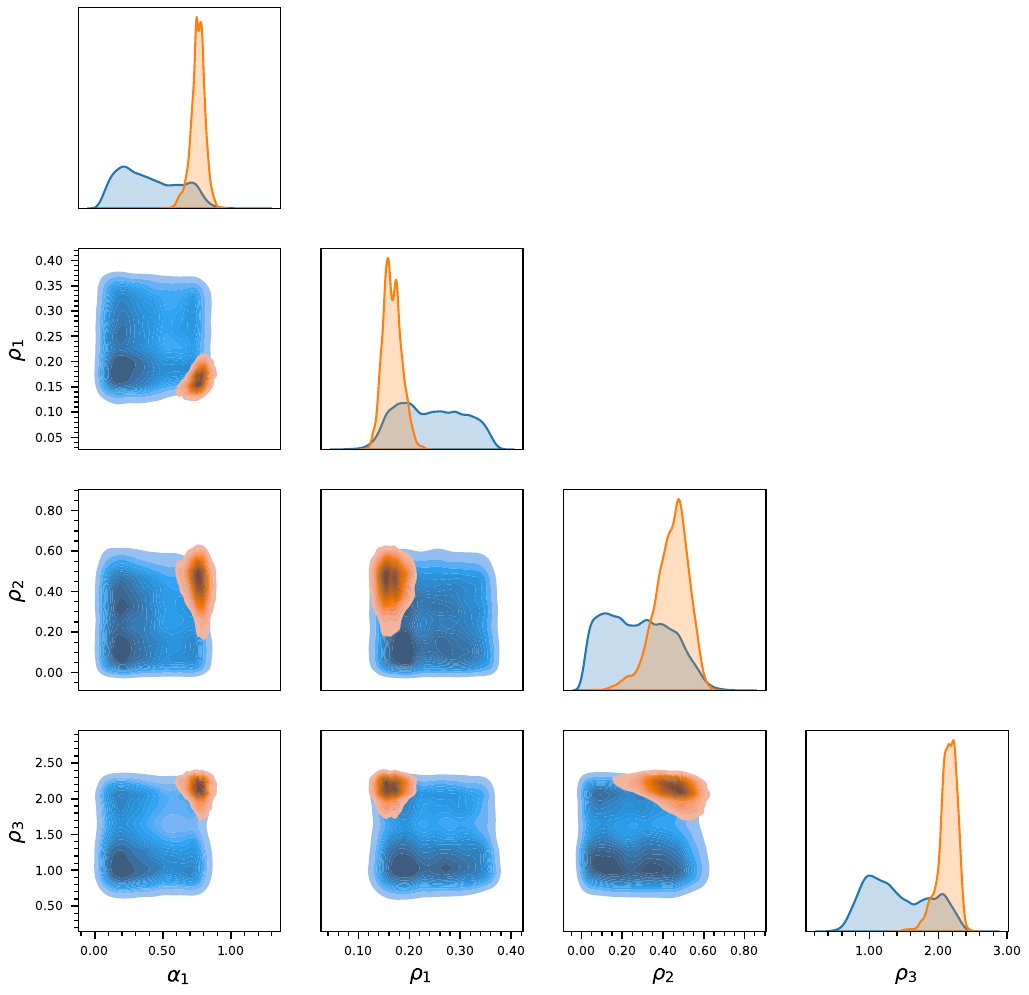}
    \end{subfigure}
    \caption{Comparison of kernel density estimates for the random samples (blue) and the MLS-recommended parameter points (orange) during the first (left) and fifth (right) iterations of the sampling scheme.}
    \label{fig:small_comp}
\end{figure}

The range of the recommended intervals for each parameter can be seen to decrease between the first and final iterations.  This behaviour is demonstrated more clearly in Fig.\ref{fig:iteration_comp} and Fig.\ref{fig:small_comp}.  The former demonstrates that the range generally decreases with each iteration.  This can be attributed to the NN gaining a `better understanding' of the parameter space, and correspondingly identifying a tighter interval where the best estimates can be found.  The lambda-type couplings in particular are seen to not have fully converged by the fifth iteration, indicating potential headroom for improvement in the recommendations.  Fig.\ref{fig:small_comp} relays similar information, showing the kernel density estimates of the parameter space across the $\alpha_1$, $\rho_1$, $\rho_2$, and $\rho_3$ couplings for both the random and recommended data sets.  This demonstrates the decreasing region of interest as the iterations proceed, and in particular demonstrates how much the modal parameter points deviate between the random and recommendation datasets.  Functionally, Fig.\ref{fig:small_comp} also serves to concretely identify the region of interest for gravitational wave observation for the left-right phase transition, a result not yet provided in studies of this model.  It demonstrates only a small fraction of the total available parameter space is accessible to probe via gravitational wave observations, and provides a suitable prior for future studies on parameter space reconstruction within the context of gravitational wave astronomy.

In Appendix \ref{App::Extra}, the corner plots seen in Fig.\ref{fig:small_comp} are shown for the full parameter space in Fig.\ref{fig:corner_1} and Fig.\ref{fig:corner_5}.  As expected from the previous results, these show little distinguishing ability between the random samples and the recommended samples for the $\alpha_{2,3}$, $\rho_4$, and $\lambda_{2,3,4}$ couplings.  This indicates the strong degeneracy of the gravitational wave signal across these parameters.  That is, the gravitational wave signal likely shows little variation when these parameters are changed.  This will pose a significant hurdle for gravitational wave inference for the LRSM.  Performing a parameter space reconstruction from gravitational wave observations using, e.g., Fischer matrix or a Bayesian analysis might prove difficult to fully determine the specific parameters.  Alternatively, such an analysis might only successfully reconstruct the $\alpha_1$ and $\rho_{1,2,3}$ couplings with any degree of certainty.

\begin{figure}[htbp]
    \centering
    \begin{subfigure}[b]{0.49\textwidth}
        \centering
        \includegraphics[width=\textwidth]{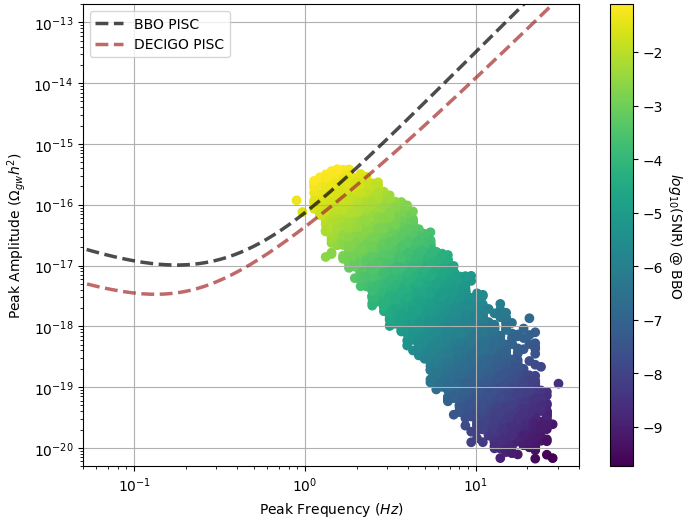}
    \end{subfigure}
    \hfill
    \begin{subfigure}[b]{0.49\textwidth}
        \centering
        \includegraphics[width=\textwidth]{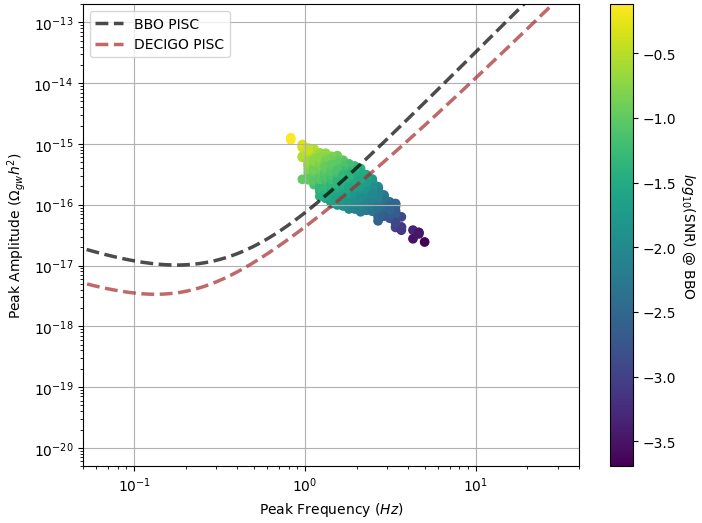}
    \end{subfigure}
    \caption{Comparison of random samples (left) to MLS-recommended samples (right) across the gravitational wave prediction space for each final data set. Each data set contains 157,093 and 33,866 parameter points respectively.}
    \label{fig:full_comp}
\end{figure}

\begin{figure}[htbp]
    \centering
    \includegraphics[width=0.6\linewidth]{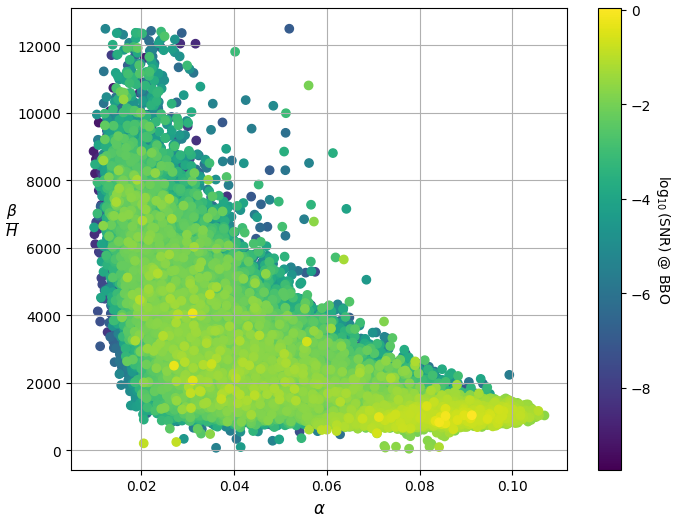}
    \caption{The SNR at BBO compared to the thermal parameters $\alpha$ and $\beta/H$.}
    \label{fig:therm}
\end{figure}

The set of random and NN recommended parameters were both provided to \texttt{PhaseTracer} to identify the true gravitational wave predictions.  A comparison of the predictions between the random and recommended samples is given in Fig.\ref{fig:full_comp}. Improvement can be seen from two sources.  Further random sampling has produced parameter points within the detectable region, whilst the neural network has correctly identified regions with the strongest gravitational amplitudes.  Furthermore, the NN has identified parts of the gravitational wave prediction space that have not yet been identified by the random sampling process.  Given the results in the previous plots, it is likely that such parts would not have been identified at all.  The dependence of the SNR on the thermal parameters $\alpha$ and $\beta/H$ is shown in Fig.\ref{fig:therm}.  In general, the transition displays a low strength $\alpha < 0.1$ and favours a large $\beta/H$.  Both conditions correspond to low detectability. In Appendix \ref{App::Extra}, the dependence of these thermal quantities on the parameters are shown in Fig.\ref{fig:alpha} and Fig.\ref{fig:beta}.  These plots show little correlation between the regions of high $\alpha$ and low $\beta/H$.  Additionally, we find that $\alpha$ is strongly correlated with high $\Omega h^2$, while $\beta/H$ is not. 

In Appendix \ref{App::Extra}, a kernel density estimate comparison of the predictions for both the random and recommended sampling is provided in Fig.\ref{fig:kde}.  This demonstrates the increased sampling efficiency of the MLS method.  Additionally, it should be clear that whilst the random sampling method can produce detectable results, it is only the MLS method that is capable of generating such results at a reasonable frequency.

\begin{figure}[htbp]
    \centering
    \includegraphics[width=0.6\linewidth]{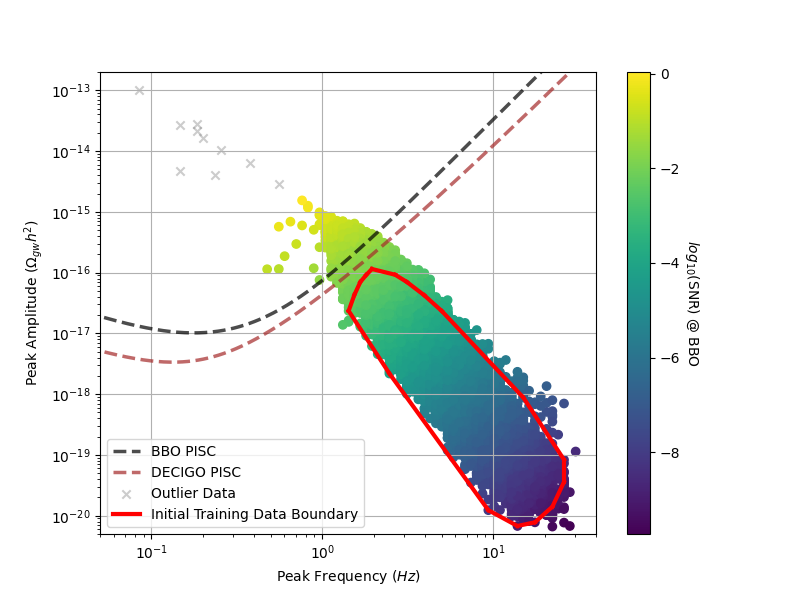}
    \caption{Final results post-MLS improvement. Clear improvements have been made to the detectability of this phase transition at near future detectors. Shown is the outline of the preliminary training data, indicating the enhancements provided by both increasing the number of random samples and adding the MLS improvement. Also shown are parameter points that have been identified as numerical outliers.}
    \label{fig:final_data}
\end{figure}

The gravitational wave predictions for the full set of parameters, combining the randomly generated and the NN recommended parameter values, are shown in Fig.\ref{fig:final_data}.  Also shown are several outlier data points not shown on previous plots, as well as the boundary of the initial training data.  This clearly demonstrates that an order-of-magnitude improvement has been obtained via an application of the MLS method.  Significantly, this improvement has resulted in parameter points that now lie within the detectable region at several near-future gravitational-wave telescopes.  The outlier data points shown here were checked to be the result of some numerical artefact by adding some small perturbation to the parameter point and then recalculating the resulting spectrum.  In all cases, the resultant amplitude fell into the $10^{-15}$ to $10^{-16}$ region.  This instability and lack of robustness strongly suggest that these points should not be taken into serious consideration when assessing the detectability of this phase transition.

Given the limited sample size we have investigated, with $\mathcal{O}(10^6)$ processed samples across a 10-dimensional parameter space, we refrain from conclusively stating that the MLS has fully explored the parameter space for this model.  However, comparing the blue regions of Fig.\ref{fig:small_comp} demonstrates little variation in the domain of the parameter space across each iteration, suggesting that the allowable parameter space is near exhaustion.  Regardless, these results serve to demonstrate how effectively the MLS scheme has succeeded in identifying the region of interest.  With as little as $\mathcal{O}(10^4)$ total samples, and $\mathcal{O}(10^3)$ samples per iteration, the scheme has found the parameter space that is detectable and consequently probable at near future observatories.

\begin{figure}[htbp]
    \centering
    \includegraphics[width=0.6\linewidth]{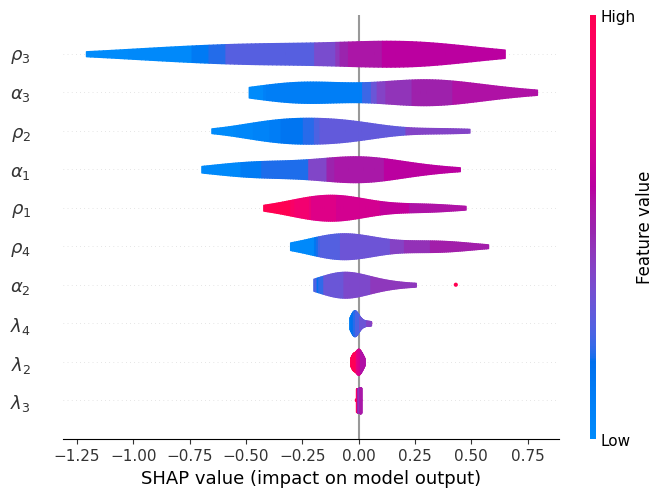}
    \caption{SHAP values of the neural network trained on the final full dataset after MLS improvement. These values indicate the predictive influence of each quartic coupling on the gravitational wave amplitude. Among them, $\rho_3$ exhibits the strongest relationship with the amplitude. In contrast, the $\lambda$-type couplings display significant degeneracy, with the model output showing minimal sensitivity to their variation.}
    \label{fig:shap}
\end{figure}

Given the final data set, a neural network was trained with the same model architecture as the MLS model to predict the peak amplitude and peak frequency of gravitational waves.  The SHAP values for each feature of this final NN are given in Fig.\ref{fig:shap}.  This serves to demonstrate how strongly the gravitational wave amplitude depends on each input quartic coupling, ordered by how strong the given relationship is.  This confirms the previous claims made of Fig.\ref{fig:corner_1} and Fig.\ref{fig:corner_5}, that is, the output depends very weakly, if at all, on the $\lambda$ type couplings.  The strongest predictor is $\rho_3$.  From Fig.\ref{fig:small_comp} (equivalently, Fig.\ref{fig:corner_1} and Fig.\ref{fig:corner_5}) the strongest gravitational wave predictions are those with a large $\rho_3$.  This suggests that the tightest restriction on deriving stronger gravitational wave predictions is the inability to test larger $\rho_3$ values, due to the presence of the Landau pole emerging for such large values.

\section{Conclusions}
\label{sec::conclusion}


In this work, we investigated the prospects of detecting a first-order phase transition in the minimal left-right symmetric model at future gravitational wave observatories such as BBO and DECIGO.  Building upon previous studies, we introduced both theoretical and computational improvements.  Using the 3dEFT framework, we provided a more accurate one-loop effective potential and incorporated updated constraints on the Higgs sector.  
While much of the parameter space yields gravitational wave signals below the sensitivity thresholds of near-future detectors, our Machine Learning Scan identifies a viable region where the predicted signals are within reach of proposed observatories such as BBO and DECIGO.  This highlights the importance of targeted searches in complex BSM landscapes.

Specifically, we find that typical transitions exhibit low strength ($\alpha \leq 0.1$) and high nucleation rates ($\beta/H \gtrsim \mathcal{O}(10^3)$), leading to suppressed gravitational wave amplitudes.  However, a small region of the parameter space yields a clearly detectable signal, with a signal-to-noise ratio in the range $SNR \sim 1$–$10$.

By leveraging the computational efficiency of \texttt{PhaseTracer}, particularly in scanning high-dimensional parameter spaces, we were able to conduct a more detailed exploration of the LRSM parameter space than in previous studies.  This enabled the identification of regions potentially accessible to gravitational wave observations.  Our results corroborate earlier findings that strong signals are associated with low values of $\rho_1$, while detectable transitions tend to favour large values of $\alpha_1$, $\rho_2$, and $\rho_3$.  Additionally, we observe a significant degeneracy among the $\lambda$-type couplings, indicating that gravitational wave measurements alone offer limited discriminative power in this sector. 
To efficiently explore the parameter space and identify regions of interest, we employed a novel application of the Machine Learning Scan (MLS) method, where deep neural networks are trained to model gravitational wave amplitudes and prioritise regions of the parameter space with the most promising signals. By focusing computational effort on promising regions, the MLS method not only accelerates the analysis but makes it feasible to discover rare configurations with observable gravitational wave signals.


Our results suggest that although first-order transitions in the minimal LRSM are typically weak, a clearly identified region of the parameter space produces gravitational wave signals within reach of proposed near-future detectors, providing concrete targets for experimental searches.

\acknowledgments

CB acknowledges support from the Australian Research Council through projects DP210101636 and LE210100015.
WS was supported by an Australian Government Research Training Program Scholarship. 
YZ is supported by NNSFC No. 12105248, No. 12335005, and by the Henan Postdoctoral Science Foundation No. HN2024003.

\newpage

\bibliographystyle{JHEP}
\bibliography{references}

\appendix

\section*{Appendix}
\addcontentsline{aoc}{section}{Appendix}

\section{Vacuum Alignment}
\label{App::VEV}
\renewcommand\theequation{\Alph{section}.\arabic{equation}}

Below is a short derivation of the desirable vacuum alignment for the three scalar fields. Due to the unbroken $U(1)_{em}$ symmetry, the VEV alignment can only have non-zero components in electrically neutral fields, leading to:
\begin{equation}
    \langle \phi \rangle = \frac{1}{\sqrt{2}} \begin{pmatrix} \kappa_1 e^{i \theta_1} & 0 \\ 0 & \kappa_2 e^{i \theta_2} \end{pmatrix}, \quad \langle \Delta_L \rangle = \frac{1}{\sqrt{2}} \begin{pmatrix} 0 & 0 \\ v_L e^{i \theta_L} & 0 \end{pmatrix}, \quad \langle \Delta_R \rangle = \frac{1}{\sqrt{2}} \begin{pmatrix} 0 & 0 \\ v_R e^{i \theta_R} & 0 \end{pmatrix}.
\end{equation}
This introduces eight dynamical fields, $\kappa_1$, $\kappa_2$, $v_L$, and $v_R$, as well as their respective phases. The gauge freedom provided by the $U(1)_{B-L}$ factor was already fixed in the definition of $\mathcal{P}$. However, we can further eliminate two of the remaining phases by fixing the $SU(2)_L$ and $SU(2)_R$ gauges. We consider the action of each VEV under the transformations
\begin{equation}
    U_L = \begin{pmatrix} e^{i \omega_L} & 0 \\ 0 & e^{- i \omega_L} \end{pmatrix} \text{ and } U_R = \begin{pmatrix} e^{i \omega_R} & 0 \\ 0 & e^{- i \omega_R} \end{pmatrix},
\end{equation}
under which the phases transform according to:
\begin{align}
    \theta_1 & \rightarrow \theta_1 + \omega_L - \omega_R, \\
    \theta_2 & \rightarrow \theta_2 - \omega_L + \omega_R, \\
    \theta_L & \rightarrow \theta_L - 2 \omega_L, \\
    \theta_R & \rightarrow \theta_R - 2 \omega_R.
\end{align}
From this, two natural choices arise. The first fixes $\omega_R = \theta_R/2$ and $\omega_L = \theta_R/2 - \theta_1$, removing the phases $\theta_1$ and $\theta_R$. The second fixes $\omega_R = \theta_R/2$ and $\omega_L = \theta_L/2$, thereby eliminating both the triplet phases $\theta_L$ and $\theta_R$. We adopt the first gauge, such that $\kappa_1$ and $v_R$ are purely real.

To proceed further, we first assume there is no CP violation in the scalar potential, such that the complex coupling $\alpha_2$ is assumed to be purely real. Then, we investigate the EWSB conditions for the fields $\{ \kappa_1, \kappa_2, v_L, v_R, \theta_2, \theta_L \}$. For $v_L, \theta_2$, and $\theta_L$, we obtain:
\begin{align}
    0 & = v_L (\alpha_1 \kappa_1^2 + (\alpha_1 + \alpha_3) \kappa_2^2 + 2 \mu_3^2 + v_R^2 \rho_3) + 4 v_L \alpha_2 \kappa_1 \kappa_2 \cos(\theta_2) \nonumber \\ 
    & \quad \quad + v_R \beta_1 \kappa_1 \kappa_2 \cos(\theta_2 - \theta_L) + v_R (\beta_2 \kappa_1^2 + \beta_3 \kappa_2^2) \cos(\theta_L) \label{eq:vLcond} \\ 
    0 & = 2 ( (v_L^2 + v_R^2) \alpha_2 + (\kappa_1^2 + \kappa_2^2) \lambda_4 + 2 \mu_2^2) \sin(\theta_2) \nonumber \\ 
    & \quad \quad + 2 \kappa_1 \kappa_2 ( 2 \lambda_2 + \lambda_3) \sin( 2 \theta_2) + v_L v_R \beta_1 \sin(\theta_2 - \theta_L)   \label{eq:theta2cond} \\
    0 & = \beta_1 \kappa_1 \kappa_2 \sin(\theta_2 - \theta_L) - (\beta_2 \kappa_1^2 + \beta_3 \kappa_2^2) \sin(\theta_L)
    \label{eq:thetaLcond}
\end{align}
From this, we immediately notice Eq.\eqref{eq:thetaLcond} is trivial if $\beta_i = 0$. Under these conditions, Eq.\eqref{eq:theta2cond} implies $\sin(\theta_2) = \sin(2 \theta_2) = 0$, from which we conclude $\theta_2 = 0$. Finally, substituting both $\beta_i = 0$ and $\theta_2 = 0$ into Eq.\eqref{eq:vLcond} allows us to immediately conclude $v_L = 0$. Therefore, assuming no CP violation and neglecting the $\beta$ type couplings in the Higgs sector, we conclude all VEVs are real, and the $\Delta_L$ VEV can be neglected entirely.

Under these assumptions, we adopt the following VEV alignment:
\begin{equation}
    \langle \phi \rangle = \frac{1}{\sqrt{2}} \begin{pmatrix} \kappa_1 & 0 \\ 0 & \kappa_2 \end{pmatrix}, \  \langle \Delta_L \rangle = 0, \ \langle \Delta_R \rangle = \frac{1}{\sqrt{2}} \begin{pmatrix} 0 & 0 \\ v_R & 0 \end{pmatrix}.
\end{equation}

\section{Scalar Eigenstates}
\label{App::Eigens}
\renewcommand\theequation{\Alph{section}.\arabic{equation}}

Here we derive the physical scalar eigenstates for the VEV alignments given in Section.~\ref{eq:VEValign}. The scalar fields are given by:
\begin{equation}
    \phi = \begin{pmatrix} \phi_1^0 & \phi_1^+ \\ \phi_2^{-} & \phi_2^0\end{pmatrix}, \quad \Delta_L = \begin{pmatrix} \delta_L^{+}/\sqrt{2} & \delta_L^{++} \\ \delta_L^0 & - \delta_L^{+}/\sqrt{2} \end{pmatrix}, \quad \Delta_R = \begin{pmatrix} \delta_R^{+}/\sqrt{2} & \delta_R^{++} \\ \delta_R^0 & - \delta_R^{+}/\sqrt{2} \end{pmatrix}.
\end{equation}
The eigenstates can be most easily derived in the following bases: $\text{Re} \{ \phi_1^0, \phi_2^0, \delta_R^0, \delta_L^0\}$, \\ $\text{Im} \{ \phi_1^0, \phi_2^0, \delta_R^0, \delta_L^0\}$, $\{ \phi_1^+, \phi_2^-, \delta_R^+, \delta_L^+\}$, and $\{ \delta_R^{++}, \delta_L^{++} \}$, corresponding to CP-even, CP-odd, singly-charged, and doubly-charged physical states. The full analytic diagonalisation is cumbersome, and as such we assume a $\kappa_2 \gg \kappa_1$ and perform the diagonalisation to zeroth order in $r = \kappa_2/\kappa_1$, neglecting terms proportional to $\kappa_1$. Furthermore, our symmetry breaking conditions have been applied to eliminate $\mu_i^2$ parameters.

\paragraph{CP Odd} The CP-odd neutral scalars, $\text{Im} \{ \phi_1^0, \phi_2^0, \delta_R^0, \delta_L^0\}$, have the mass matrix:
\begin{equation}
    M^2 = \begin{pmatrix}
        0 & 0 & 0 & 0 \\
        0 & \frac{1}{2} \alpha_3 v_R^2 & 0 & 0 \\
        0 & 0 & 0 & 0 \\
        0 & 0 & 0 & \frac{1}{2}(\rho_3 - 2 \rho_1) v_R^2
    \end{pmatrix}.
\end{equation}
From this, we identify two goldstone bosons corresponding to the degrees of freedom for the left and right handed $Z$ bosons, together with two massive CP-neutral scalars $A_1^0$ and $A_2^0$ with masses:
\begin{equation}
    m^2(A_1^0) = \frac{1}{2}(\rho_3 - 2 \rho_1) v_R^2, \quad m_{A_2^0}^2 = \frac{1}{2}  \alpha_3 v_R^2.
\end{equation}

\paragraph{Singly Charged} The singly charged mass matrix, corresponding to $\{ \phi_1^+, \phi_2^-, \delta_R^+, \delta_L^+\}$, is given by:
\begin{equation}
    M^2 = \begin{pmatrix}
        \frac{1}{2} \alpha_3 v_r^2 & 0 & \frac{1}{2 \sqrt{2}} \alpha_3 \kappa_1 v_R & 0 \\
        0 & 0 & 0 & 0 \\
        \frac{1}{2 \sqrt{2}} \alpha_3 \kappa_1 v_R & 0 & \frac{1}{4} \alpha_3 \kappa_1^2 & 0 \\
        0 & 0 & 0 & \frac{1}{4} \alpha_3 \kappa_1^2 + \frac{1}{2} (\rho_3 - 2 \rho_1 ) v_R^2 
    \end{pmatrix}.
\end{equation}
From which we again identify two null eigenvalues corresponding to four singly charged goldstone boson degrees of freedom, which are given to the left and right handed $W$ gauge boson masses. We then have the singly charged scalar particles, $H^{\pm}_1$ and $H^{\pm}_2$, with masses given by:
\begin{equation}
    m^2(H_1^{\pm}) = \frac{1}{4} ( \alpha_3 \kappa_1^2 - 4 v_R^2 \rho_1 + 2 v_R^2 \rho_3), \quad m^2(H^{\pm}_2) = \frac{1}{4} \alpha_3 (2 v_R^2 + \kappa_1^2).
\end{equation}
\paragraph{Doubly Charged} The doubly charged matrix corresponds to the basis $\{ \delta_R^{++}, \delta_L^{++} \}$ and is given by:
\begin{equation}
    M^2 = \begin{pmatrix}
        \frac{1}{2} \alpha_3 \kappa_1^2 + 2 v_R^2 ( \rho_1 + \rho_2) & 0 \\
        0 & \frac{1}{2} \alpha_3 \kappa_1^2 + \frac{1}{2} (\rho_3 - 2 \rho_1 ) v_R^2 \\
    \end{pmatrix}.
\end{equation}
From which we identify the masses of the doubly charged scalars $H^{\pm \pm}_1$ and $H^{\pm \pm}_2$ as:
\begin{equation}
    m^2(H_1^{\pm \pm}) = \frac{1}{2} \alpha_3 \kappa_1^2 + \frac{1}{2} (\rho_3 - 2 \rho_1 ) v_R^2, \quad m^2(H^{\pm \pm}_2) = \frac{1}{2} \alpha_3 \kappa_1^2 + 2 v_R^2 ( \rho_1 + \rho_2).
\end{equation}
\paragraph{CP-even} The mass matrix for the CP-even scalars, corresponding to $\text{Re} \{ \phi_1^0, \phi_2^0, \delta_R^0, \delta_L^0\}$, is given by:
\begin{equation}
    \begin{split}
    M^2 & = \begin{pmatrix}
        2 \lambda \kappa_1^2 & 2 \lambda_4 \kappa_1^2 & \alpha_1 v_R \kappa_1 & 0 \\
        2 \lambda_4 \kappa_1^2 & \frac{1}{2} \alpha_3 v_R^2 + 2 (\lambda_2 + \lambda_3) \kappa_1^2 & 2 \alpha_2 v_R \kappa_1 & 0 \\
        \alpha_1 v_R \kappa_1  & 2 \alpha_2 v_R \kappa_1 & 2 \rho_1 v_R^2 & 0 \\
        0 & 0 & 0 & \frac{1}{2} (\rho_3 - 2 \rho_1 ) v_R^2 \\
    \end{pmatrix} \\ & = v_R^2 \left (  \begin{pmatrix}
        0 & 0 & 0 & 0 \\
        0 & \frac{1}{2} \alpha_3 & 0 & 0 \\
        0 & 0 & 2 \rho_1 & 0 \\
        0 & 0 & 0 & \frac{1}{2} (\rho_3 - 2 \rho_1 ) \\
    \end{pmatrix} + \epsilon \begin{pmatrix}
        0 & 0 & \alpha_1 & 0 \\
        0 & 0 & 2 \alpha_2 & 0 \\
        \alpha_1 & 2 \alpha_2 & 0 & 0 \\
        0 & 0 & 0 & 0 \\
    \end{pmatrix} + \epsilon^2 \begin{pmatrix}
        2 \lambda_1 & 2 \lambda_4 & 0 & 0 \\
        2 \lambda_4 & 2 \lambda_3 + 4 \lambda_2 & 0 & 0 \\
        0 & 0 & 0 & 0 \\
        0 & 0 & 0 & 0 \\
    \end{pmatrix} \right ).
    \end{split}
\end{equation}
In the second equality, we have written the mass matrix in the form $M^2_{ij} = v_R^2 ( D_{ij} + \epsilon A_{ij} + \epsilon^2 B_{ij} )$, where $\epsilon = \kappa_1 / v_R \ll 1$. The eigenstates are then obtained perturbatively in $\epsilon$. To second order, these are given by:
\begin{equation}
    m^2_{i} = v_R^2 \bigg (D_{ii} + \epsilon^2 \sum_{i \ne j} \frac{A_{ij}}{D_{ii} - D_{jj}} + \epsilon^2 B_{ii} \bigg )
\end{equation}
Note there is no $\mathcal{O}(\epsilon)$ contribution as $A_{ii} = 0 \ \forall i$. From this, we obtain the eigenstates for our CP-even scalars $H_{0, 1, 2, 3}^0$ as:
\begin{align}
    m^2(H_0^0) & = \bigg( 2 \lambda_1 - \frac{\alpha_1}{2 \rho_1} \bigg ) \kappa_1^2 \label{eq:PhysHiggsMass} \\
    m^2(H_1^0) & = \frac{1}{2} \alpha_3 v_R^2 + 2 \bigg ( 2 \lambda_2 + \lambda_3 + \frac{2 \alpha_2}{ \alpha_3 - 4 \rho_1} \bigg ) \kappa_1^2 \\ 
    m^2(H_2^0) & = 2 \rho_1 v_R^2 + \bigg ( \frac{\alpha_1}{2} - \frac{4 \alpha_2}{\alpha_3 - 4 \rho_1 } \bigg ) \kappa_1^2\\ 
    m^2(H_3^0) & = \frac{1}{2} (\rho_3 - 2 \rho_1) v_R^2 \\ 
\end{align}
The only physical scalar without a mass of the order $\mathcal{O}(v_R)$ is $H_0^0$, which we identify as the SM Higgs, $h$, and will be used to match the quartic coupling $\lambda_1$.

\section{Gravitational Wave Fitting Functions}
\label{App::GW}
\renewcommand\theequation{\Alph{section}.\arabic{equation}}

For reference, here we provide a collection of gravitational wave results. For bubble collisions, the relevant formulas are:
\begin{align}
    \mathcal{A}_{\text{col}}(v_w) & = 1.67 \times 10^{-5} \times \frac{0.48 v^3}{1 + 5.3v_w^2 + 5 v_w^4}, \\
    n_{\beta, \text{col}} & = 2, \\
    n_{\alpha, \text{col}} & = 2, \\
    \mathcal{S}_{\text{col}} (x) & = \frac{1}{0.064 x^{-3} + 0.456 x^{-1} + 0.48x)} \\
    \frac{f_{\text{peak}, \text{col}}}{\mu Hz} & = 16.5 \left ( \frac{f_\star}{\beta} \right ) \left (\frac{g_\star}{100} \right )^{1/6} \left( \frac{\beta_\star}{H_\star} \right ) \left ( \frac{T_\star}{100 \text{ GeV} } \right ).
\end{align}
Here $x = f/f_{\text{peak}}$ and we have:
\begin{equation}
    \frac{f_\star}{\beta} = \frac{0.35}{1 + 0.069 v_w + 0.69 v_w^4}.
\end{equation}
The efficiency factor is given by:
\begin{equation}
    \kappa_{\text{col}} = \frac{1}{1 + 0.517\alpha} \left ( 0.715 \alpha + \frac{4}{27} \sqrt{\frac{3 \alpha }{2}} \right ).
\end{equation}
For gravitational waves generated by MHD turbulence, we have:
\begin{align}
    \mathcal{A}_{\text{turb}}(v_w) & = 3.35 \times 10^{-4} v_w, \\
    n_{\beta, \text{turb}} & = 1, \\
    n_{\alpha, \text{turb}} & = 3/2, \\
    \mathcal{S}_{\text{col}} (x) & = \frac{x^3}{(1 + x)^{11/3} ( 1 + 8 \pi f/H_0)}, \\
    \frac{f_{\text{peak}, \text{turb}}}{\mu Hz} & = \frac{27}{v_w} \left (\frac{g_\star}{100} \right )^{1/6} \left( \frac{\beta_\star}{H_\star} \right ) \left ( \frac{T_\star}{100 \text{ GeV} } \right ).
\end{align}
Again $x = f/f_{\text{peak}}$ and the red-shifted Hubble constant is given by:
\begin{equation}
    H_0 = 16.5 \left (\frac{g_\star}{100} \right )^{1/6}  \left ( \frac{T_\star}{100 \text{ GeV} } \right ) \mu \text{Hz{}}
\end{equation}
The efficiency factor is related to the sound wave efficiency factor as $\kappa_{\text{turb}} = 0.1 \kappa_{\text{sw}}$.\

For sound waves, the stated gravitational wave formula in is ~\cite{Athron:2024xrh}:
\begin{equation}
    \Omega_{\text{sw}} h^2 = 2.061 F_{\text{gw, 0}} (\Gamma \overline{U}_f^2)^2 \mathcal{S}_{\text{sw}}(x) \tl{\Omega}_{\text{gw}} \min(H_\star R_\star/\overline{U}_f, 1) (H_\star R_\star),
\end{equation}
where $R_\star$ is the mean bubble separation and is given by:
\begin{equation}
    R_\star = (8 \pi)^{1/3} v_w/\beta_\star.
\end{equation}
Clearly, we have $H_\star R_\star = (H_\star/\beta_\star) (8 \pi)^{1/3} v_w$. Likewise, we exchange 
\begin{equation}
    F_{\text{gw}, 0} = 3.57 \times 10^{-5} \left( \frac{g_\star}{100} \right )^{-1/3}.
\end{equation}
$\Gamma$ is the ratio of enthalpy to energy density for the fluid, taken to be $4/3$ for the early universe, whilst $\overline{U}_f$ is the enthalpy-weighted RMS fluid velocity. As such, we have:
\begin{equation}
    \Gamma \overline{U}_f^2 = \frac{\kappa_{sw} \alpha }{1 + \alpha}
\end{equation}
These definitions allow for the minimum $\min(H_\star R_\star/\overline{U}_f, 1)$ to be evaluated, and we have the numerical factor $ \tl{\Omega}_{\text{gw}} \sim 0.012$ which arises from matching to simulation data. Combining the above expressions, we arrive at the following definitions:
\begin{align}
    \mathcal{A}_{\text{sw}}(v_w) & = 2.16 \times 10^{-4} v_w, \\
    n_{\beta, \text{sw}} & = 1, \\
    n_{\alpha, \text{sw}} & = 2, \\
    \mathcal{S}_{\text{sw}} (x) & = x^3 \left (\frac{7}{4 + 3 x^2} \right)^{7/2} \\
    \frac{f_{\text{peak}, \text{sw}}}{\mu Hz} & = \frac{0.89}{v_w} \left ( \frac{z_p}{10} \right ) \left (\frac{g_\star}{100} \right )^{1/6} \left( \frac{\beta_\star}{H_\star} \right ) \left ( \frac{T_\star}{100 \text{ GeV} } \right ).
\end{align}
Here $z_p$ is determined from simulations to be $z \sim 10$. The efficiency factor of the sound wave factor can be found from in eq. (88) of ~\cite{Athron:2024xrh}.

\section{Supplementary Findings}
\label{App::Extra}
\renewcommand\theequation{\Alph{section}.\arabic{equation}}

\begin{figure}[htbp]
    \centering
    \includegraphics[width=\linewidth]{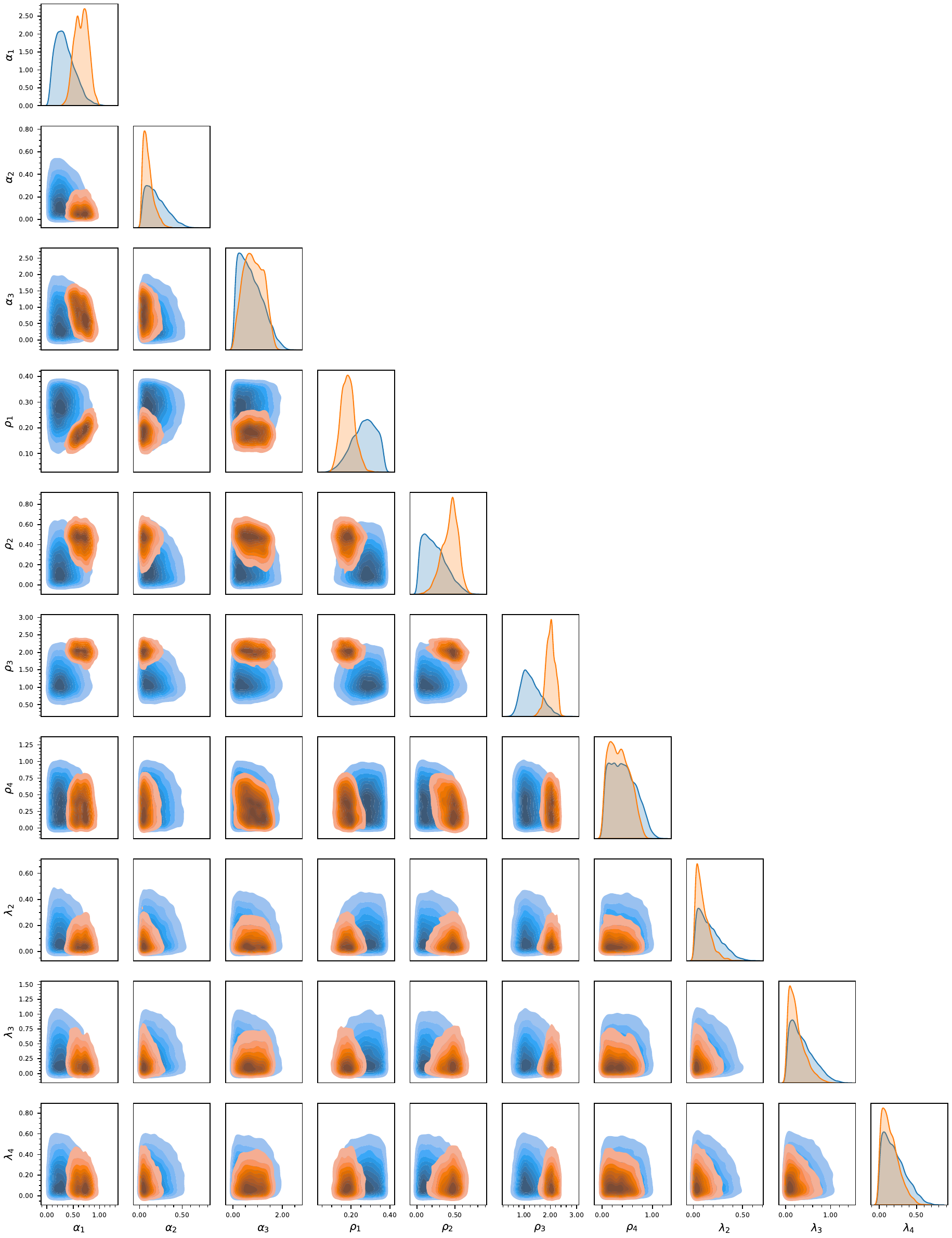}
    \caption{Kernel density estimate across the full parameter space for the random samples (blue) and recommended samples (orange) for the first iteration. There is little observed variance between the two datasets for the $\alpha_3$ and $\lambda$-type couplings.}
    \label{fig:corner_1}
\end{figure}

\begin{figure}[htbp]
    \centering
    \includegraphics[width=\linewidth]{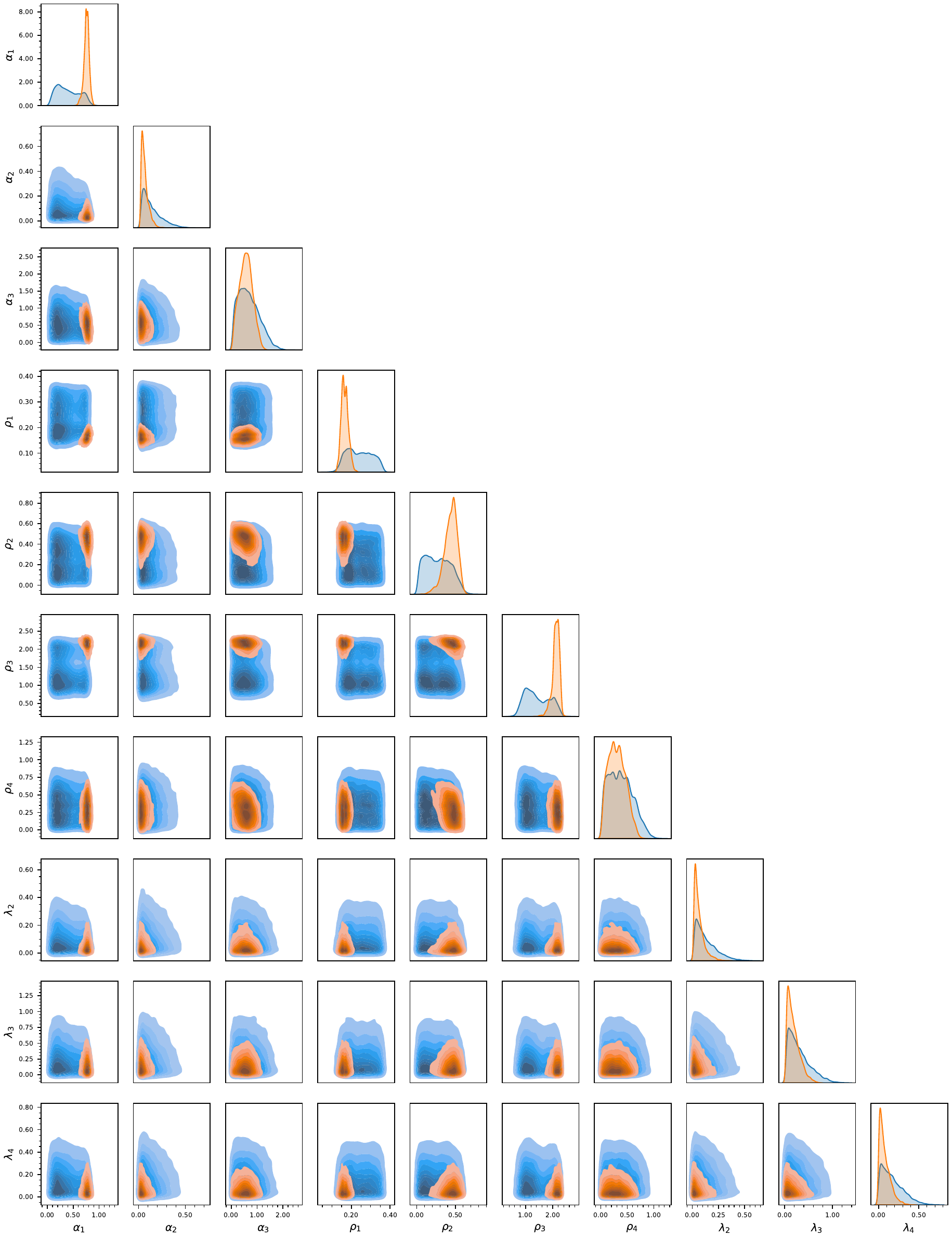}
    \caption{Kernel density estimate across the full parameter space for the random samples (blue) and recommended samples (orange) for the final iteration.}
    \label{fig:corner_5}
\end{figure}

The corner plots shown in Fig.\ref{fig:small_comp} can be extended to summarise the entire parameter space. In Fig.\ref{fig:corner_1} and Fig.\ref{fig:corner_5}, these plots are shown for the entire parameter space for the first and last MLS-iterations. In general, the recommended component of the parameter space by the NN agrees with the highest density regions of the random sampling, as seen in particular for the $\alpha_{2,3}$, $\rho_4$, and $\lambda_{2,3,4}$ couplings in the first iteration. This behaviour is expected. Many of these parameters provide a negligible contribution to the scalar eigenstates in the one loop potential. Indeed, these parameters only contribute to the potential via their appearance in the expressions for the 3d couplings during the dimensional reduction.

\begin{figure}[htbp]
    \centering
    \includegraphics[width=\linewidth]{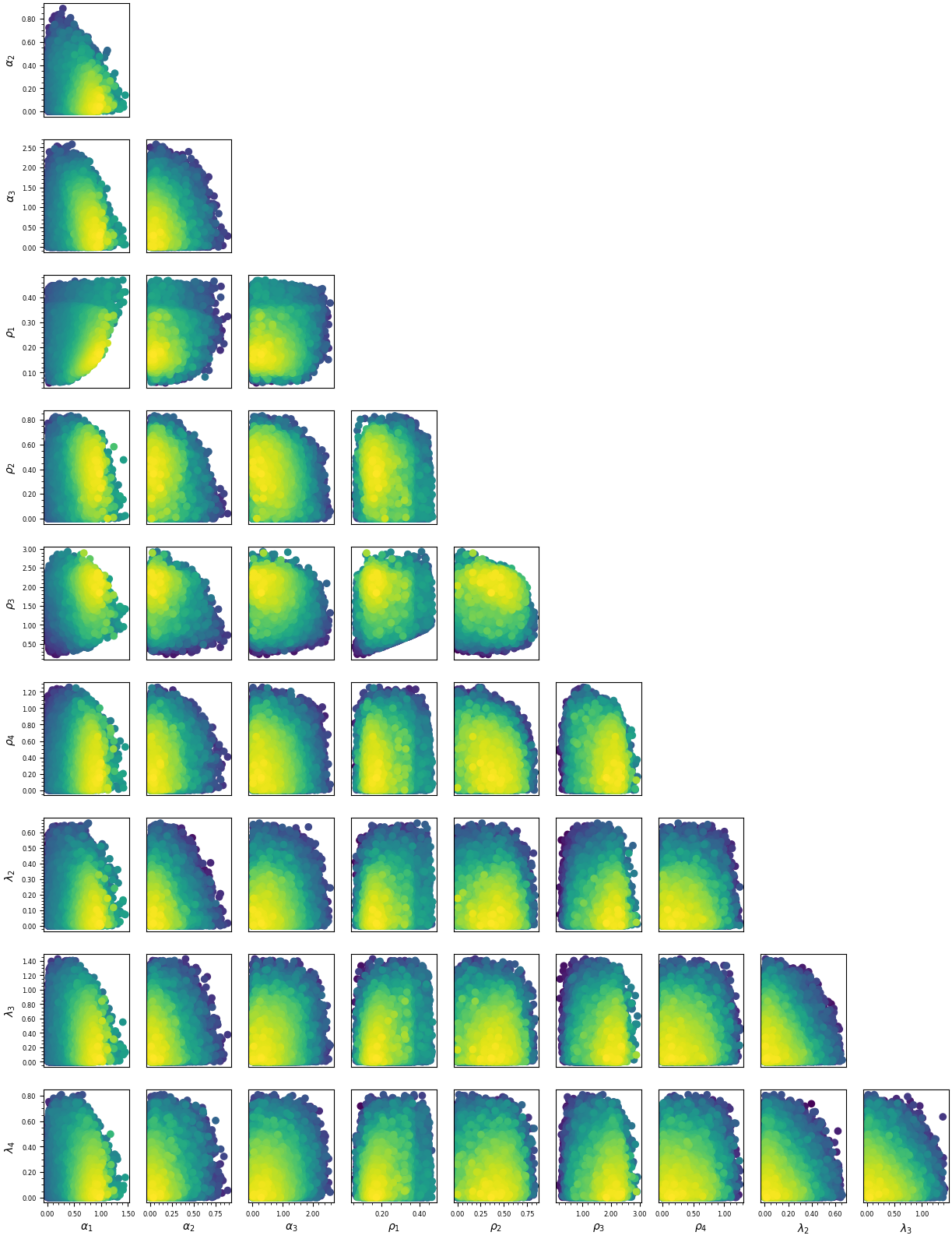}
    \caption{Variation of $\alpha$ across the LRSM parameter space.}
    \label{fig:alpha}
\end{figure}

\begin{figure}[htbp]
    \centering
    \includegraphics[width=\linewidth]{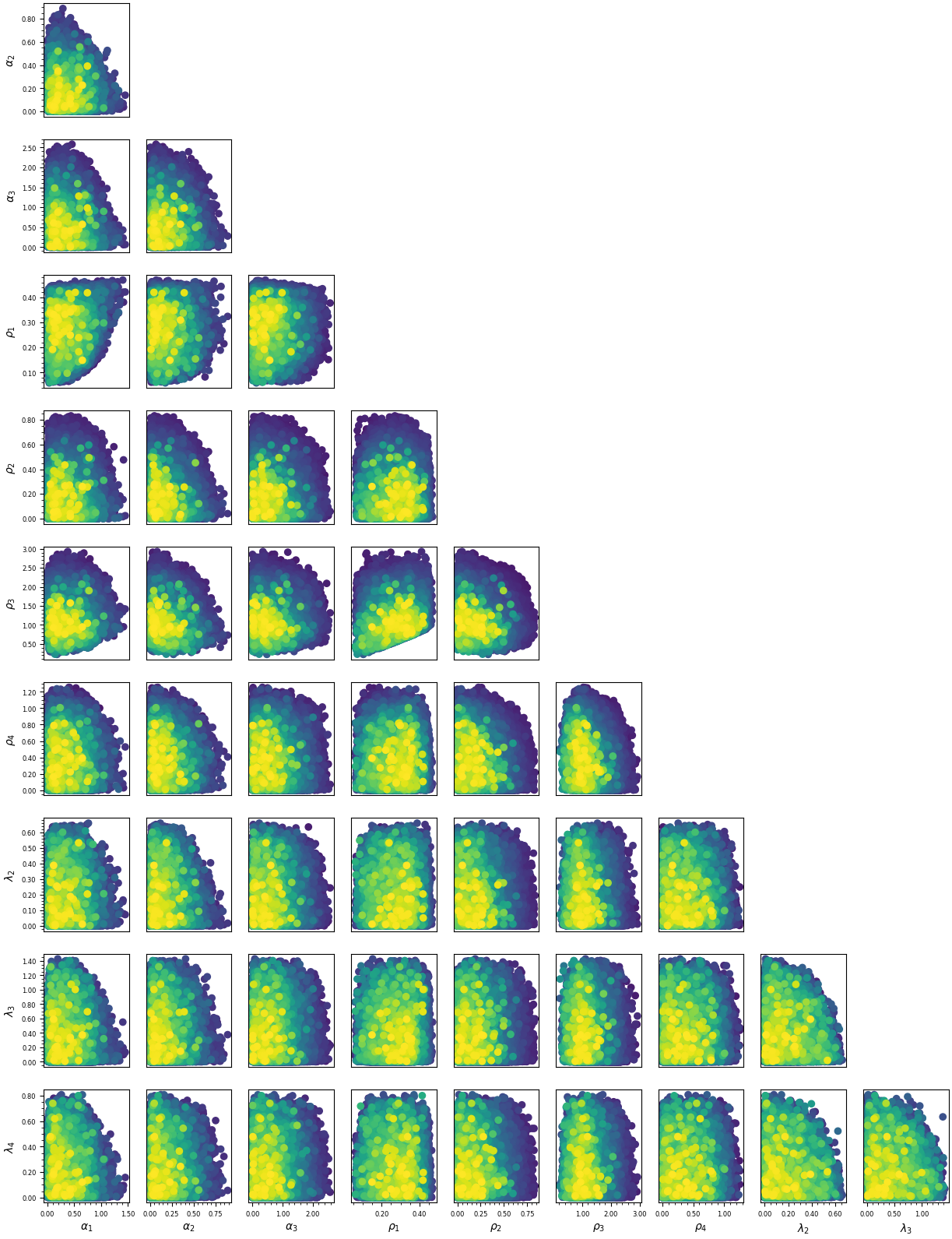}
    \caption{Variation of $\beta/H$ across the LRSM parameter space.}
    \label{fig:beta}
\end{figure}

The variation of the thermal parameters $\alpha$ and $\beta/H$ is shown across the parameter space in figures Fig.\ref{fig:alpha} and Fig.\ref{fig:beta} respectively. Note that there is not necessarily a correspondence between regions with the highest $\alpha$ and regions with the lowest $\beta/H$. Comparing these images to Fig.\ref{fig:corner_1}, we see the NN favours regions of high $\alpha$, opposed to low $\beta/H$, when determining the strongest gravitational wave estimates. There is a strong correlation between $\alpha$ and $\beta/H$ in the $\lambda_i$-type couplings, supporting the theory these play little role in generating strong predictions.

\begin{figure}[htbp]
    \centering
    \includegraphics[width=0.6\linewidth]{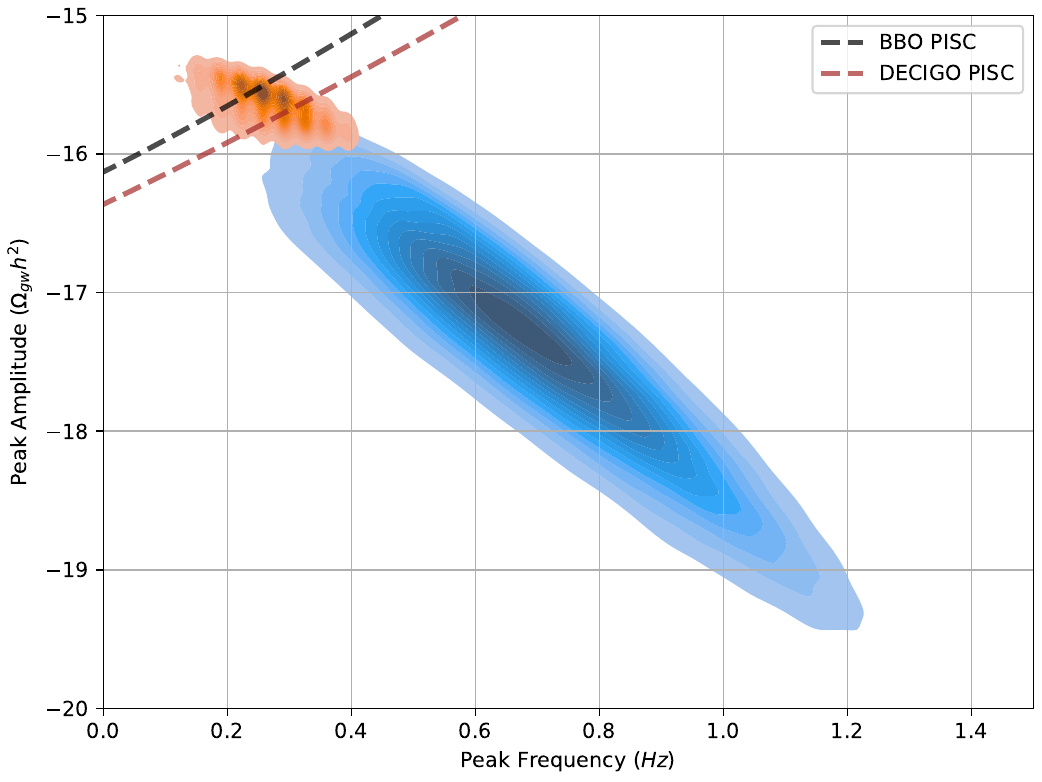}
    \caption{Kernel density estimate across the gravitational wave prediction space. This showcases that whilst both random samples and the MLS method can generate detectable signals, it is only the latter that is capable of doing so at a reasonable frequency.}
    \label{fig:kde}
\end{figure}

We showcase the improved efficiency of the MLS method in Fig.\ref{fig:kde}, demonstrating that the random sampling method is unable to provide consistently detectable parameter points.

\end{document}